\def \<{\langle}
\def \>{\rangle}

\newcommand{\ra}{\;\raise1.0pt\hbox{$'$}\hskip-6pt\partial\;}
\newcommand{\lo}{\;\overline{\raise1.0pt\hbox{$'$}\hskip-6pt\partial}\;}

\newcommand{\degree}{^\circ}

\newcommand{\Abs}{\abstract}
\newcommand{\Ack}{\acknowledgments}
\newcommand{\mktt}{\maketitle}

\documentclass[a4paper,11pt]{article}
\pdfoutput=1

\usepackage{jcappub,graphicx,epsfig,natbib,color,times,bm,amsmath,multirow,hyperref,amssymb,mathtools}
\usepackage[ddmmyyyy,hhmmss]{datetime}

\begin{document}

\title{Comparison of the Planck 2018 CMB polarization maps in the BICEP2/Keck region}

\author[a,b]{Hao Liu,}\emailAdd{liuhao@nbi.dk}

\author[a]{James Creswell}\emailAdd{james.creswell@nbi.ku.dk}

\author[a]{and Pavel Naselsky}\emailAdd{naselsky@nbi.dk}

\affiliation[a]{The Niels Bohr Institute, University of Copenhagen, Blegdamsvej 17, DK-2100 Copenhagen, Denmark}

\affiliation[b]{Key Laboratory of Particle and Astrophysics, Institute of High Energy Physics, CAS, 19B YuQuan Road, Beijing, China, 100049}

\Abs{\\
We examine the statistical properties of polarization maps from Planck 2018 within the patch of sky observed by the BICEP2/Keck experiment using the one point distribution function (1PDF), skewness, and kurtosis statistics. Our analysis is performed for the Q and U Stokes parameters and for
the corresponding E- and B-modes of the CMB signals. We extend  our analysis by studying the correlations between CMB polarization maps and residual maps (the difference between the full  signal and the CMB map) for the frequency range of 100--217 GHz with both the Q/U and
E/B approaches. Although all the CMB maps reveal almost Gaussian statistical properties for Q/U
and E/B domains, we have detected very significant anomalies for cross-correlations with
residuals at 100 GHz at the level of $3.7\sigma$ for the Commander map and $5.2\sigma$ for NILC,
for both the Q and U parameters. Using the NILC--Commander difference, which does not contain a cosmological signal, we find a sub-dominant non-Gaussian component in Q skewness and kurtosis
at the level of $4.3\sigma$ and $10\sigma$, respectively. For the B-mode we have found a very high
level of cross-correlation (0.63--0.69) between the NILC/Commander maps and the 143 GHz total signal, which cannot be associated with the cosmological component. These strong deviations suggest that remnants of foregrounds, systematic effects, and component separation exist in the 2018 Planck CMB polarization maps in the BICEP2 sky area, which is far away from the Galactic plane.  Our analysis also demonstrates the preferability of the Q/U domain over E/B for determination of the statistical properties of the derived
CMB signals, due to non-locality of the transition Q/U $\rightarrow$ E/B. 

}

\mktt

\section{Introduction}

     The next generation of CMB
experiments~\citep{2012SPIE.8442E..19H, 2016arXiv161002743A,
2011arXiv1110.2101K, quijote2012, doi.10.1093.nsr.nwy019, 2020arXiv200110272B}
is dedicated to one of the most important tasks of fundamental
physics: the detection of cosmological gravitational waves. To achieve this
goal, it is necessary to detail existing models of polarization foregrounds,
especially for the B-mode. For this it seems important to us to use
prior experience, including the BICEP2/Keck Array experiment~\cite{2015ApJ...811..126B}(hereafter, the BICEP2 zone) in combination
with Planck 2018 CMB products~\cite{Akrami:2018mcd}, in order to understand in which direction we
should focus both in the modeling of foregrounds and in the methods for extracting
the cosmological signal. In this article, we analyze the
properties of four CMB polarization maps, SMICA, NILC, Commander, and SEVEM, from the
Planck 2018 data release in the sky region of the BICEP2 experiment in two
main ways. First of all, we will be interested in the Pearson
cross-correlation coefficient~\cite{1895RSPS...58..240P} between CMB maps and
residuals (the difference between the full signal and the CMB), which should
to be at the level of chance correlations if the CMB does not contain
residuals from foregrounds and systematics. Secondly, we will study the
distribution function (1PDF, skewness and kurtosis) for CMB products in the
BICEP2 zone in order to understand how the remnants of foregrounds and
systematics deviate these distributions from Gaussian expectations. 

The non-triviality of using this combined approach is dictated by
following circumstances. Usually, Gaussian or non-Gaussian,
the derived CMB maps have specific Gaussian signs of the primordial cosmological signal and its contamination by the remnants of foregrounds and systematics \cite{Kamionkowski}. However, the BICEP2 zone is located far from the Galactic plane, and the non-Gaussianity of foregrounds is no longer their characteristic feature. As shown in \cite{Liu:2017ycx,Sebastian}, the deviation of the full-sky signal for the skewness and kurtosis statistics does not exceed 2 standard deviations for the 353 GHz sky map.
Therefore, the test for cross-correlation between the derived CMB signal and
relevant residuals becomes very important for understanding the level of contamination.

The main idea of our tests is that Planck 2018 CMB
products (SMICA, NILC, Commander and SEVEM) are
derived from the full sky analysis with different kind of masks and they based
on optimisation of different functionals under various assumptions ~\cite{2018arXiv180706205P}.
Subtraction of the mean values of these functionals makes the methods globally
non-local. We have two additional tasks among others:
\begin{enumerate}
\item To understand the properties of these polarisation CMB maps in the BICEP2
   domain, which lies far away from the Galactic masks.
\item To compare them with local ILC map~\cite{2004ApJ...612..633E}, derived just
   from BICEP2 domain for all 30--353 GHz total maps. We will perform our
   analysis  on the Q and U components of the Stokes parameters, and  look at
   the corresponding E and B components~\cite{PhysRevD.55.1830,
   1997PhRvD..55.7368K, 0004-637X-503-1-1}. We use $N_\mathrm{side}=256$ and $1^o$
   Gaussian beam smoothing, focusing on the first 100--200 multipoles.
\end{enumerate}
Working with the Q and U maps, we can apply different statistics for
amplitudes, making the corresponding statistics very informative. We will
show that tests of non-Gaussianity of the E and B maps without the Q and U components can
 produce misleading effects.

 The outline of the paper is the following. In section 2 we
will describe statistical characteristics of the Q and U parameters for the
Planck 2018 CMB maps, focusing on 1PDF, skewness, and kurtosis statistics in
the BICEP2 zone. Here we will introduce the residuals of the total signal as a
difference between  the 100--217 GHz frequency maps and the SMICA, NILC, Commander and SEVEM
maps. For comparison, we will include in the analysis our ILC map (local
ILC), derived just from the 30--353 GHz  signal in BICEP2 domain. Section 3 is
devoted to the same analysis of the statistical properties of the polarization
maps as for the Q and U domain, but for the corresponding E- and B-modes. Due to the
very small fraction of the sky occupied by the  BICEP2 zone, we have applied
recycling E/B-leakage correction, proposed in \cite{2018arXiv181104691L,
Liu_2019_EB_general}. In Section 4 we investigate the skewness and kurtosis
statistics for E- and B-modes of polarisation. We summarise our results in the
Conclusion.

\section{Planck 2018 CMB products in the BICEP2 region}\label{sec:method and
result}

\begin{figure*}[!htb]
  \centerline{
 \includegraphics[width=0.4\textwidth]{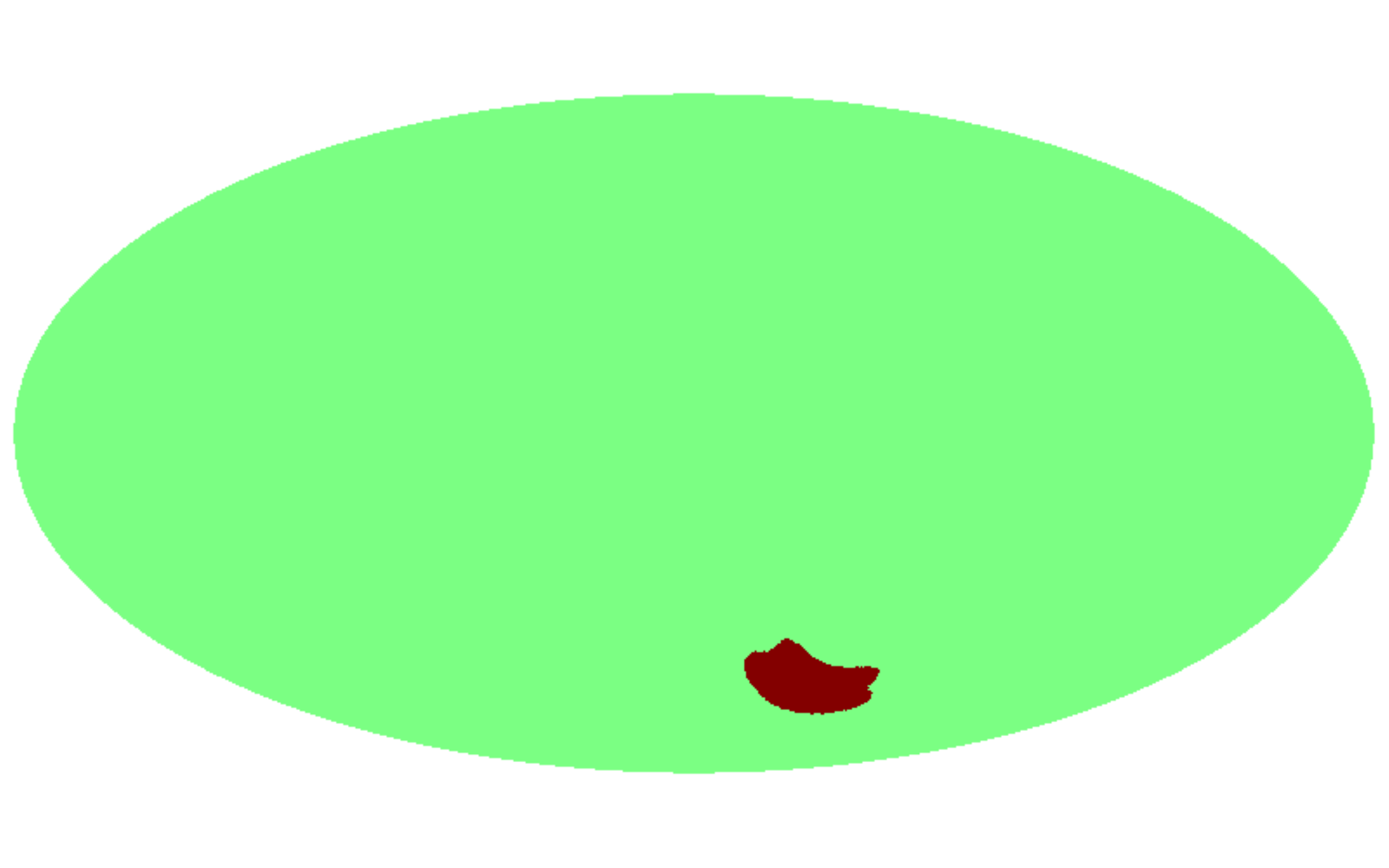} 
 \includegraphics[width=0.4\textwidth]{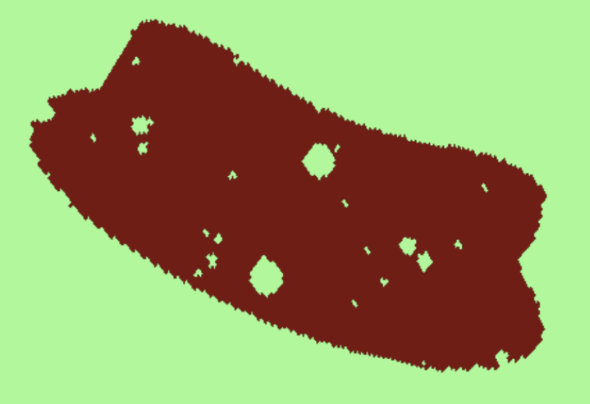}}
 \centerline{
 \includegraphics[width=0.4\textwidth]{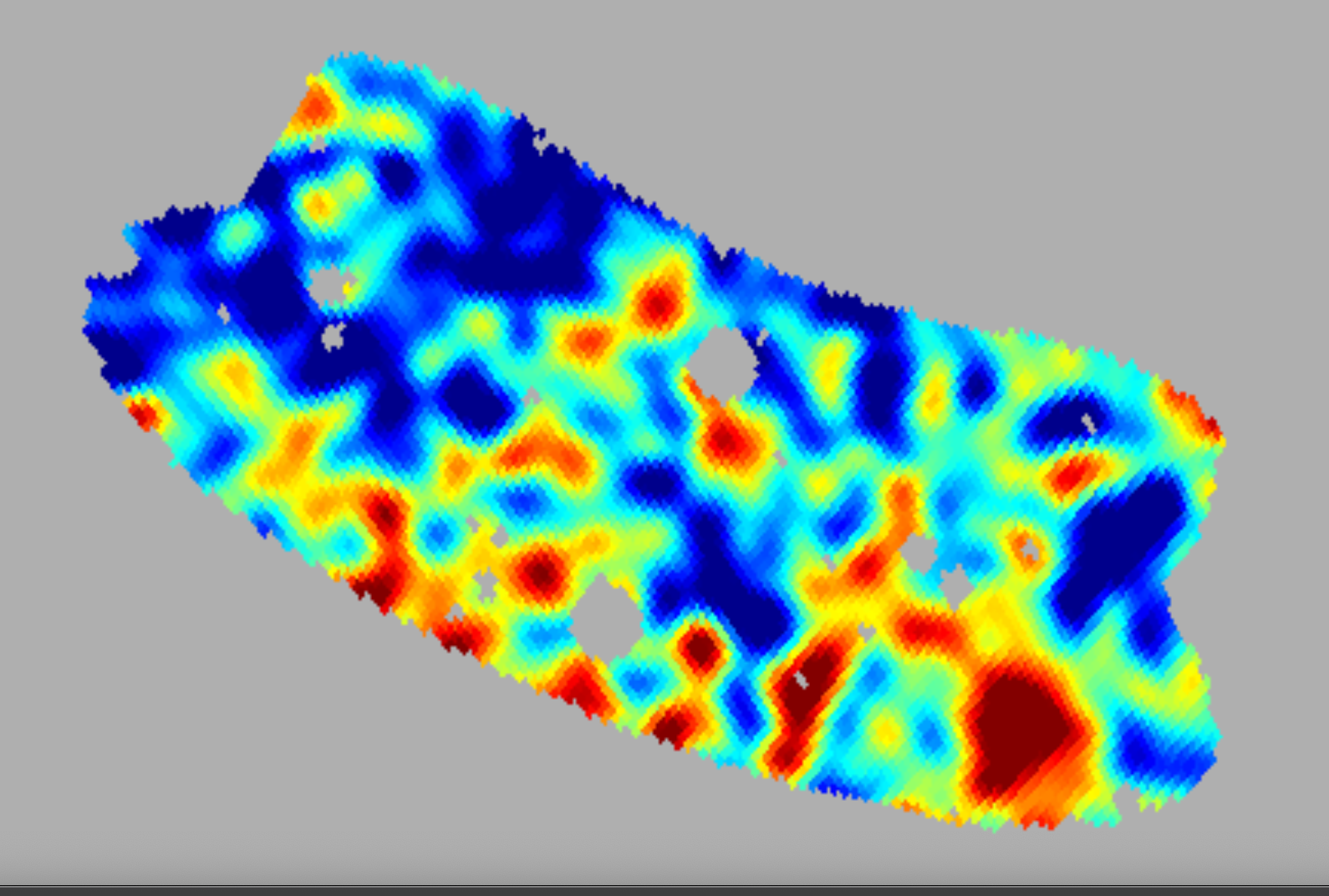}
 \includegraphics[width=0.4\textwidth]{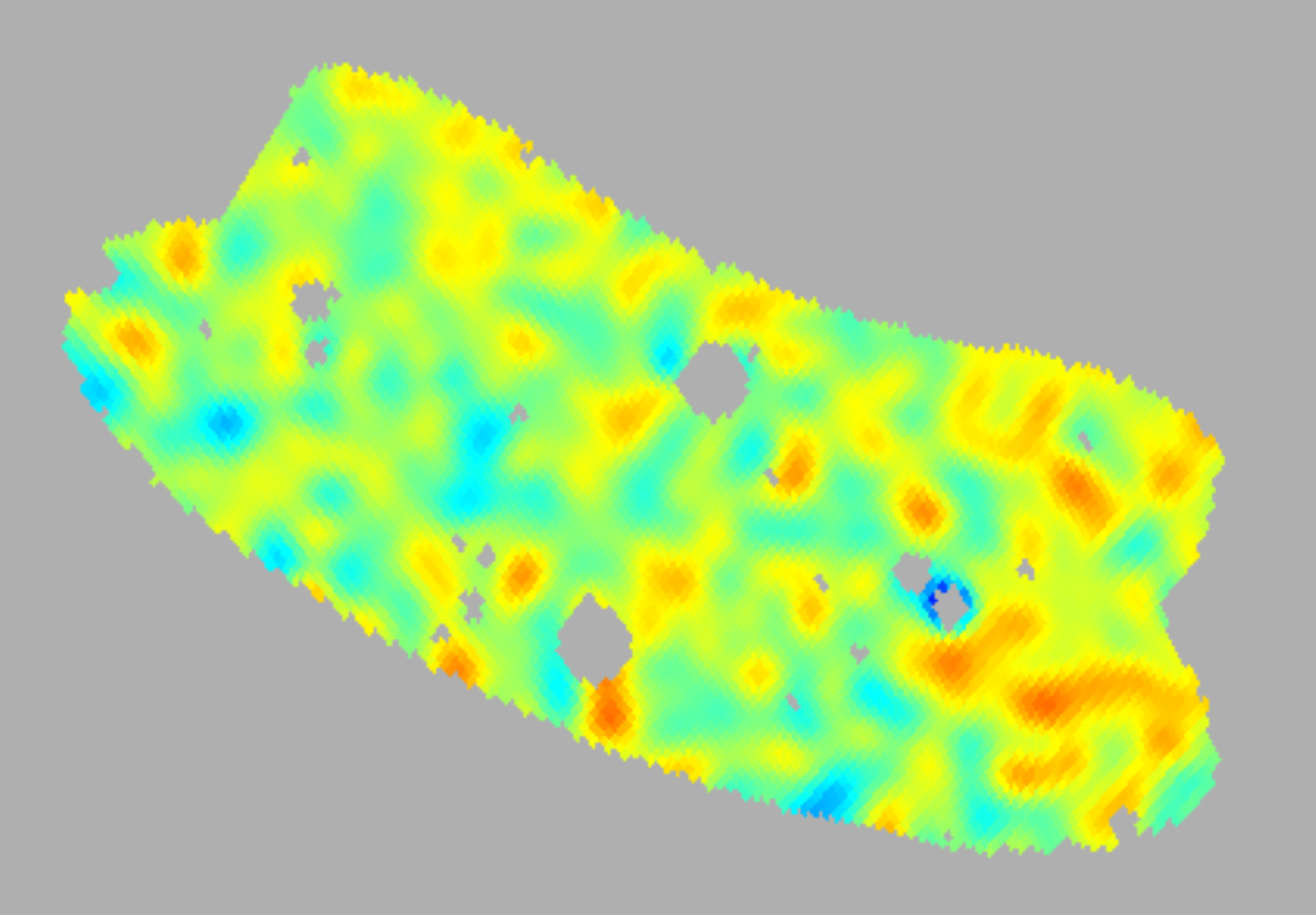}}
 \caption{Upper panels: the BICEP2 zone and the Planck point sources mask which define the
 region under investigation. Lower left: the Commander map in this region. Lower right: The difference Commander$-$NILC map. Both the lower panels are for the temperature anisotropy.
 The color scale of the maps is $-100$ to $+100 \mu$K.}
  \label{figbicep}
\end{figure*}

In this section we compare the polarized CMB maps, SMICA, NILC, Commander, and
SEVEM, from the Planck 2018 data release.
These maps are derived from a masked full sky
analysis, which we then restrict to the BICEP2 region, shown in figure~\ref{figbicep}.
We will focus on the varying statistical properties of these maps, which reflect
different methods of separating foregrounds from the primordial signal.

We will add to our analysis the ILC map derived by us from the Planck 2018
30-353 GHz maps. The ILC coefficients are computed using the convenient
solution given in~\cite{2004ApJ...612..633E}. All these maps are smoothed by a
$1\degree$ FWHM Gaussian beam.

In figure~\ref{figbicep}, the BICEP2 region is presented with implementation of
the Union mask in that region for point-like sources. For illustration of
the method we plot in that figure the temperature anisotropy maps for
Commander and the difference between Commander and NILC maps. Both these maps,
Commander and NILC, are characterised by very high signal-to-noise-ratio, and
if they are Gaussian, the difference between them represents the residuals of
the foregrounds and instrumental noise.
 
\subsection{Asymmetry of distributions}

\begin{figure*}[!htb]
  \centering
  \includegraphics[width=0.48\textwidth]{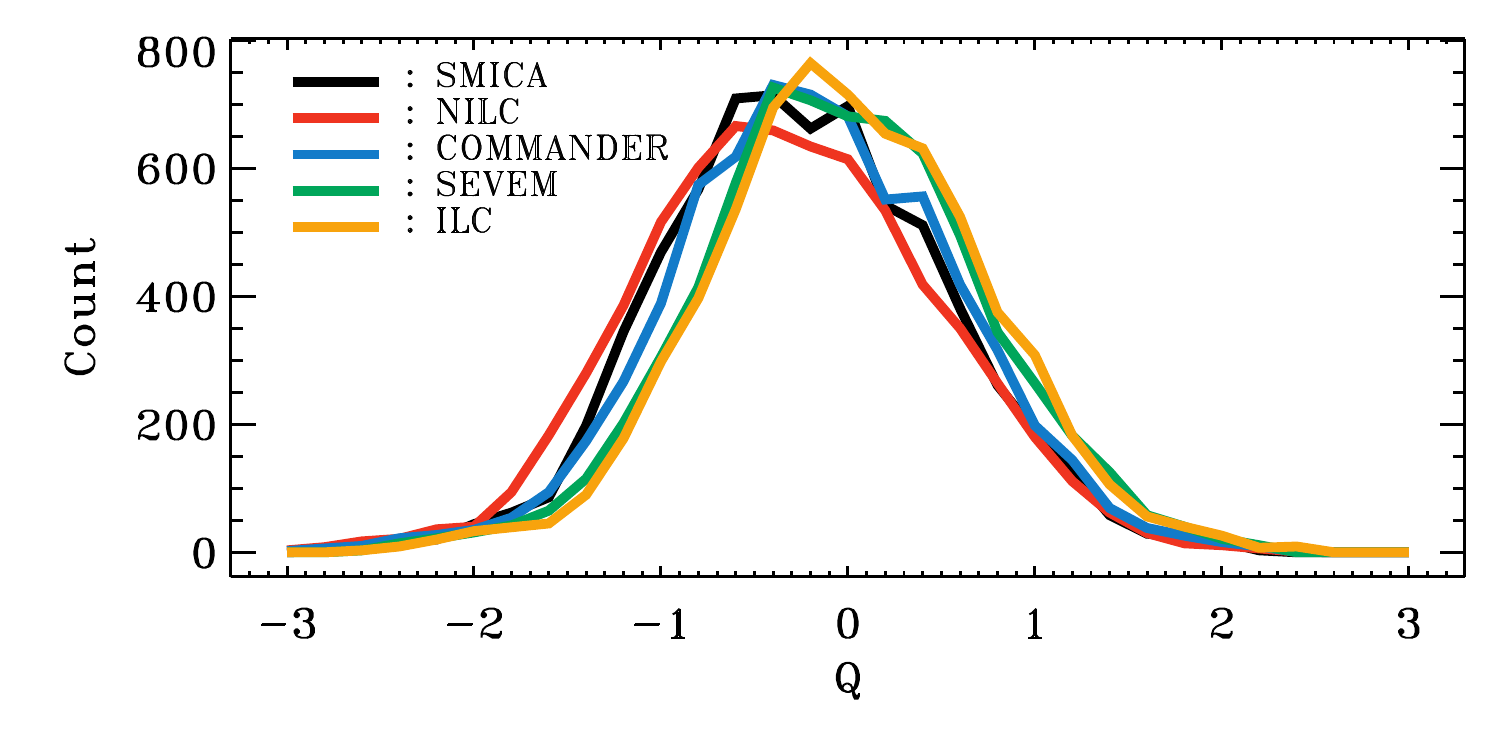}
  \includegraphics[width=0.48\textwidth]{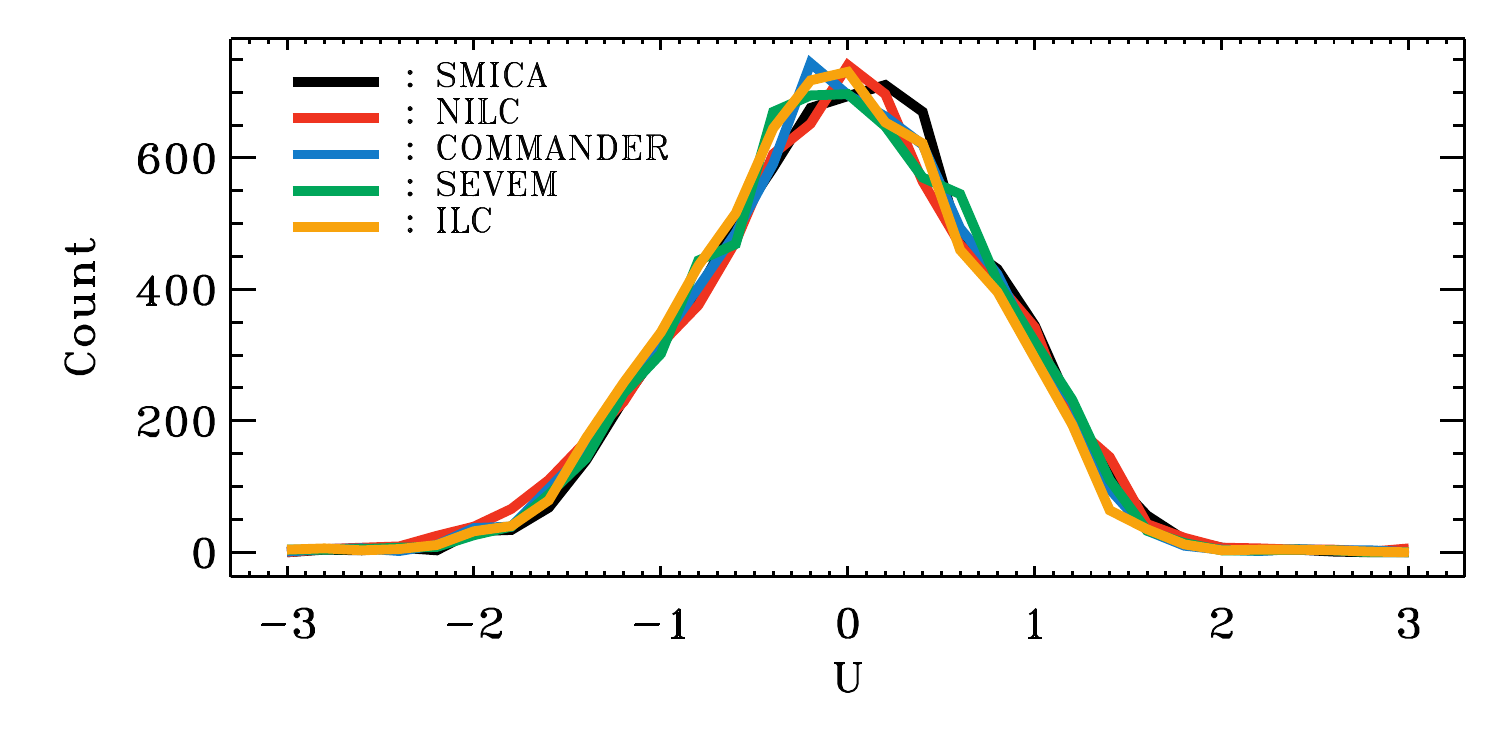}
  
  \caption{ The histograms of the Q and U Stokes parameters (left and right) of
  the regions in figure~\ref{figbicep} for the CMB maps. }
  \label{fig:hist QU cmb}
\end{figure*}

As a first step of our analysis, in figure~\ref{fig:hist QU cmb} we show the
distribution function (histogram) of counts with given amplitudes of the Q and U components for the
five maps in the BICEP2 zone. The variation of these distributions from each other is pronounced for Q but more stable for U.

The first test that can be applied for these histograms is the analysis of
skewness and kurtosis. We will characterise the departure of these
characteristics from Gaussian realisations in terms of standard deviations.
For that we will generate $10^3$ Gaussian realisations in order to estimate
the variance of distribution $\sigma$. Then in table~\ref{tablea} we present
the corresponding significance of non-Gaussian features of distribution in
terms of $\sigma$.
\begin{table}[!htb]
 \centering
 \begin{tabular}{|c|c|c|c|c|} \hline
    Input & Q skewness & Q kurtosis & U skewness & U kurtosis \\ \hline
    SMICA           & 0.7 & 0.4 & 1.8 & 0.4 \\ \hline 
    NILC            & 0.3 & 0.3 & 1.0 & 1.2 \\ \hline 
    Commander       & 0.3 & 1.4 & 1.4 & 0.3 \\ \hline 
    SEVEM           & 0.2 & 0.9 & 1.5 & 0.1 \\ \hline 
    ILC             & 0.2 & 1.4 & 1.3 & 0.5 \\ \hline 
\end{tabular}
 \caption{ The significances (multiples of $\sigma$ of estimator in the Gaussian case) of the skewness and kurtosis 
 of the Q and U Stokes parameters for various maps. }
 \label{tablea}
\end{table}
\begin{figure*}[!htb]
 \includegraphics[width=0.3\textwidth]{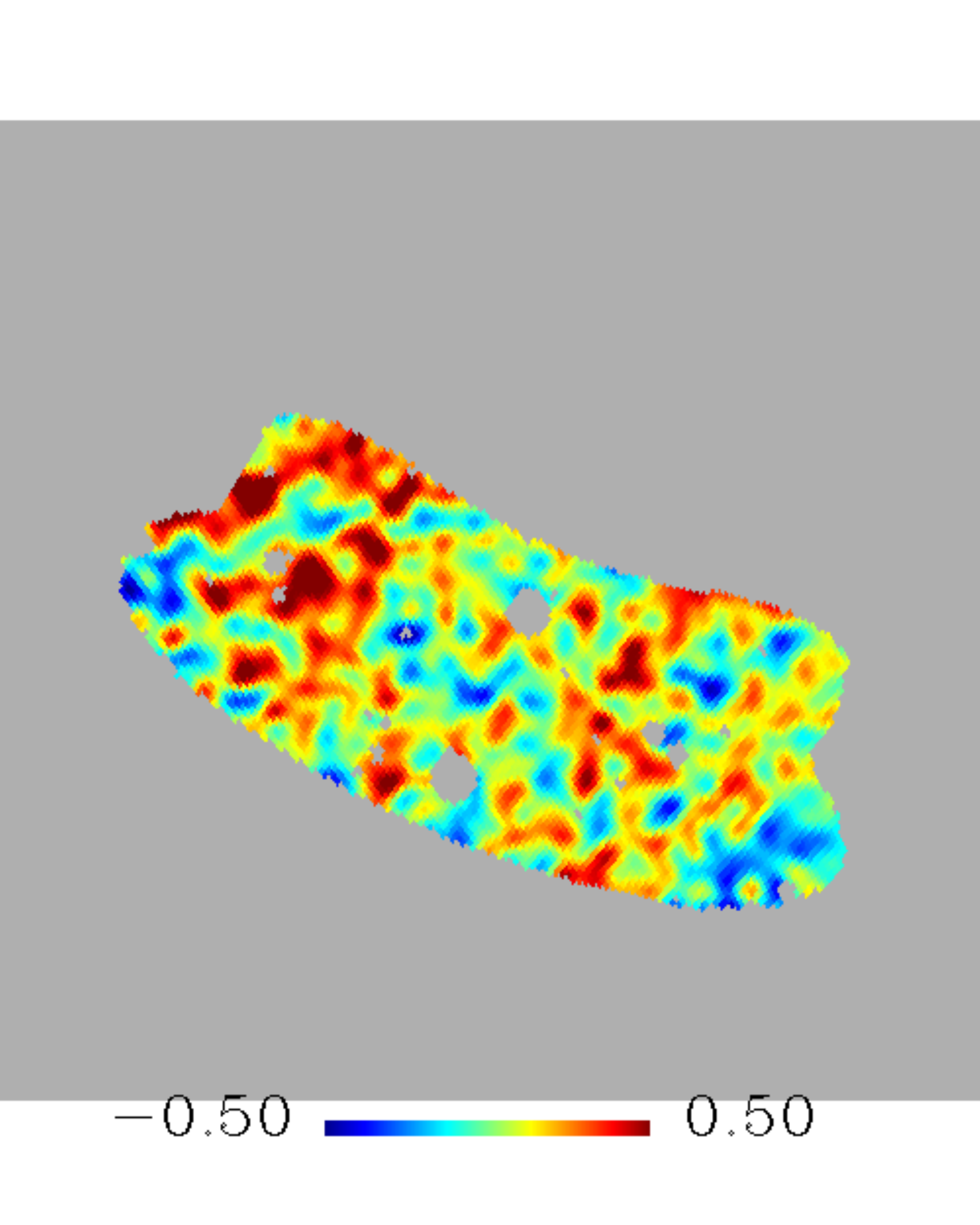}
 \includegraphics[width=0.3\textwidth]{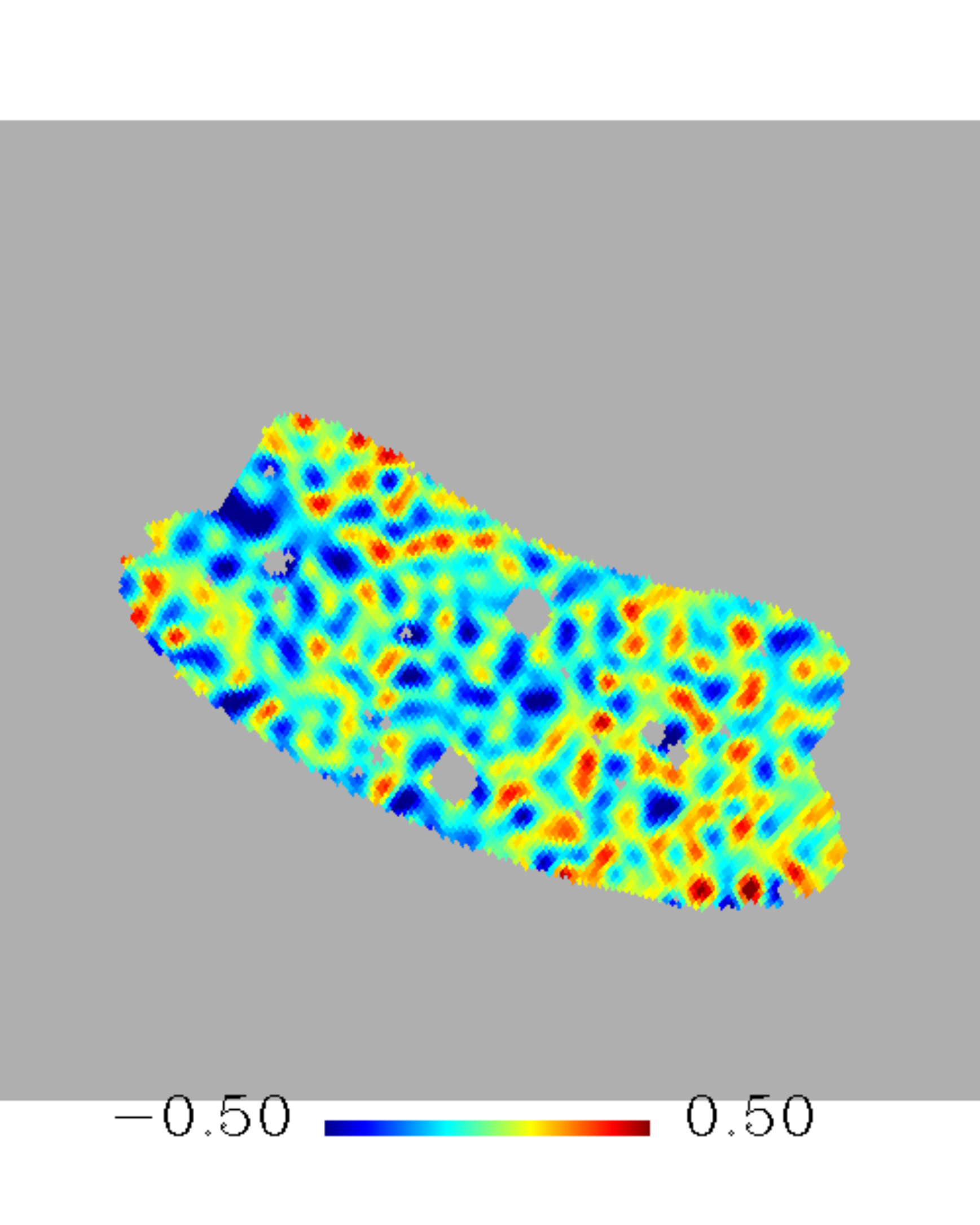}
 \includegraphics[width=0.3\textwidth]{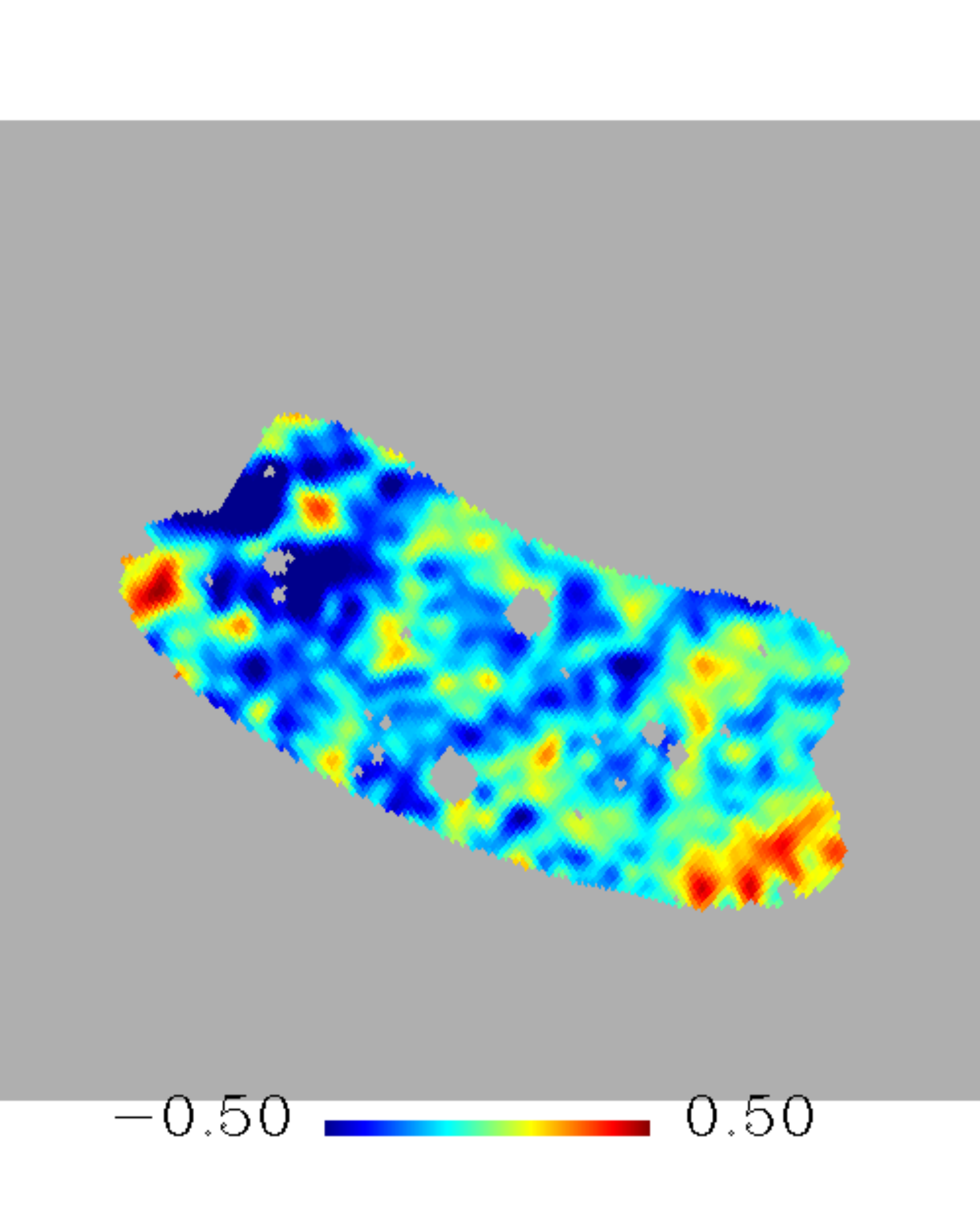}
 
 \includegraphics[width=0.3\textwidth]{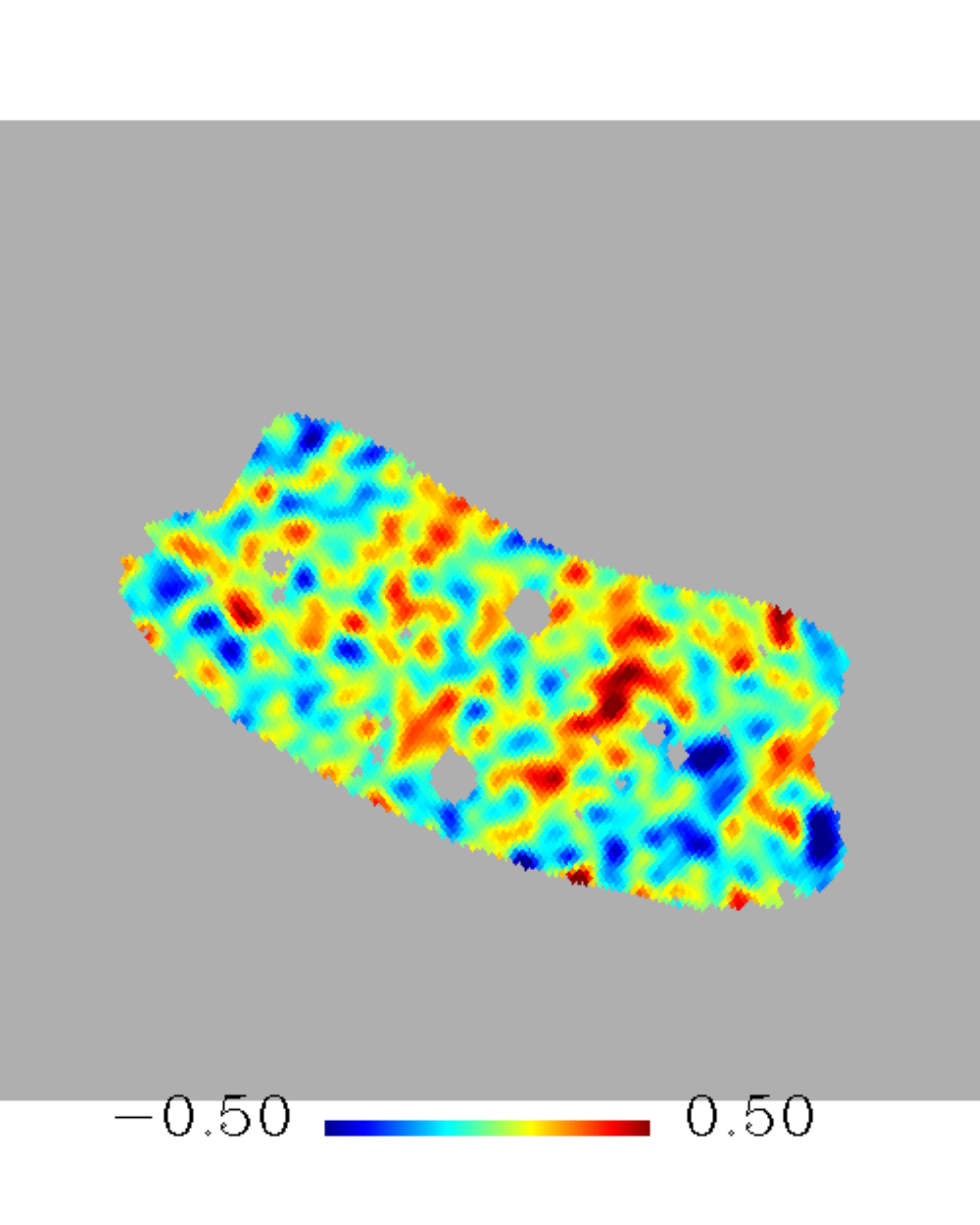}
 \includegraphics[width=0.3\textwidth]{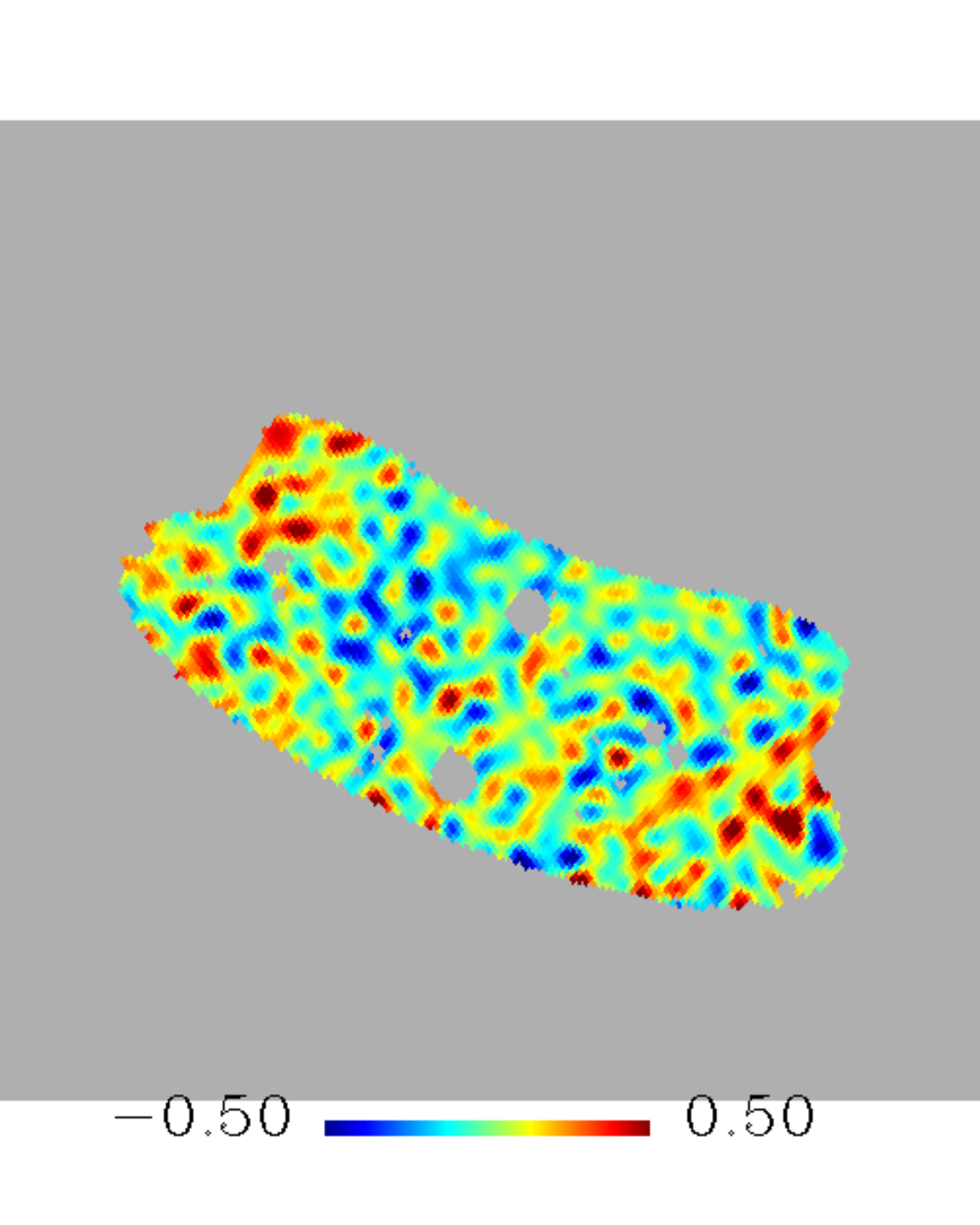}
 \includegraphics[width=0.3\textwidth]{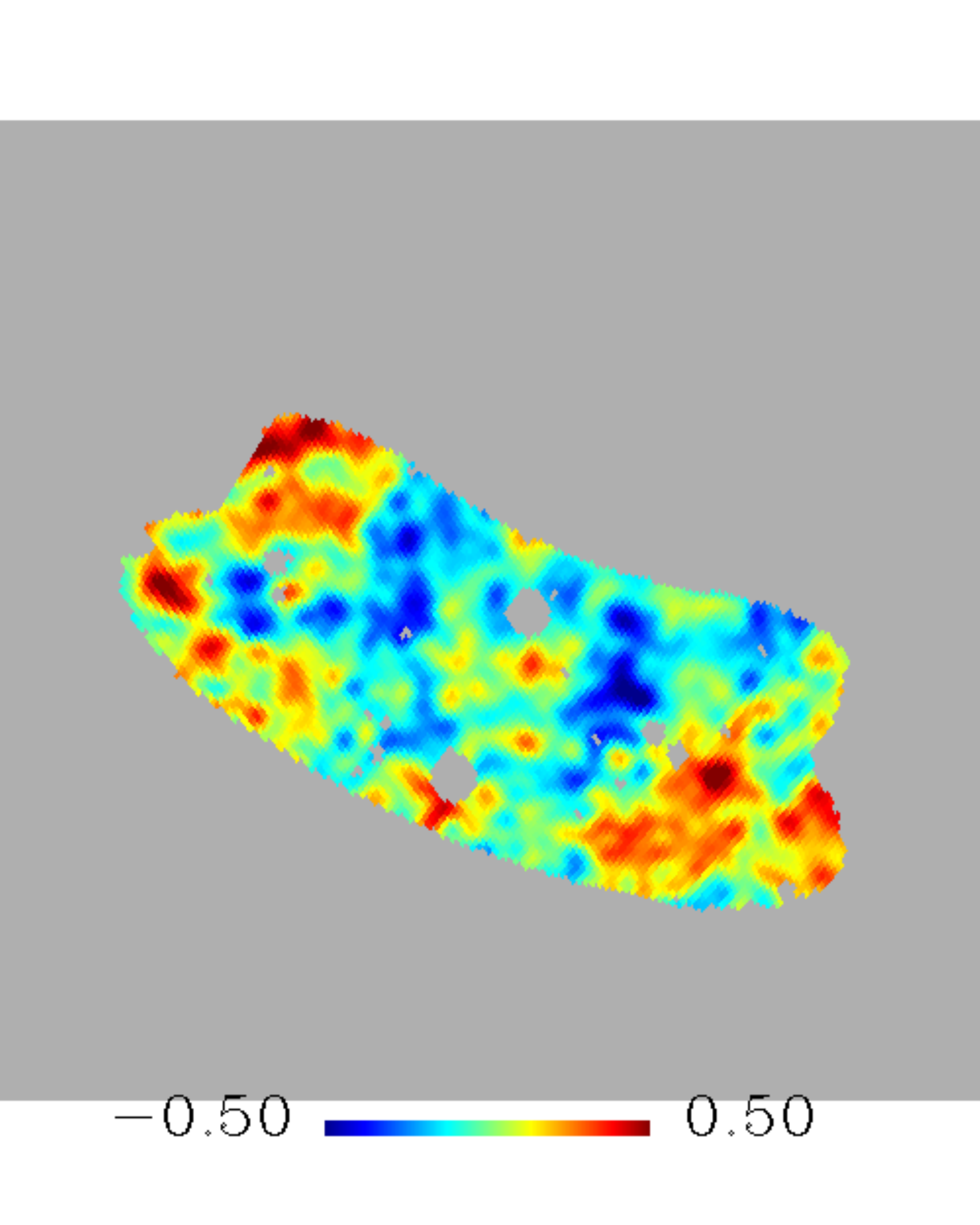}
 \centerline{ 
  \includegraphics[width=0.6\textwidth]{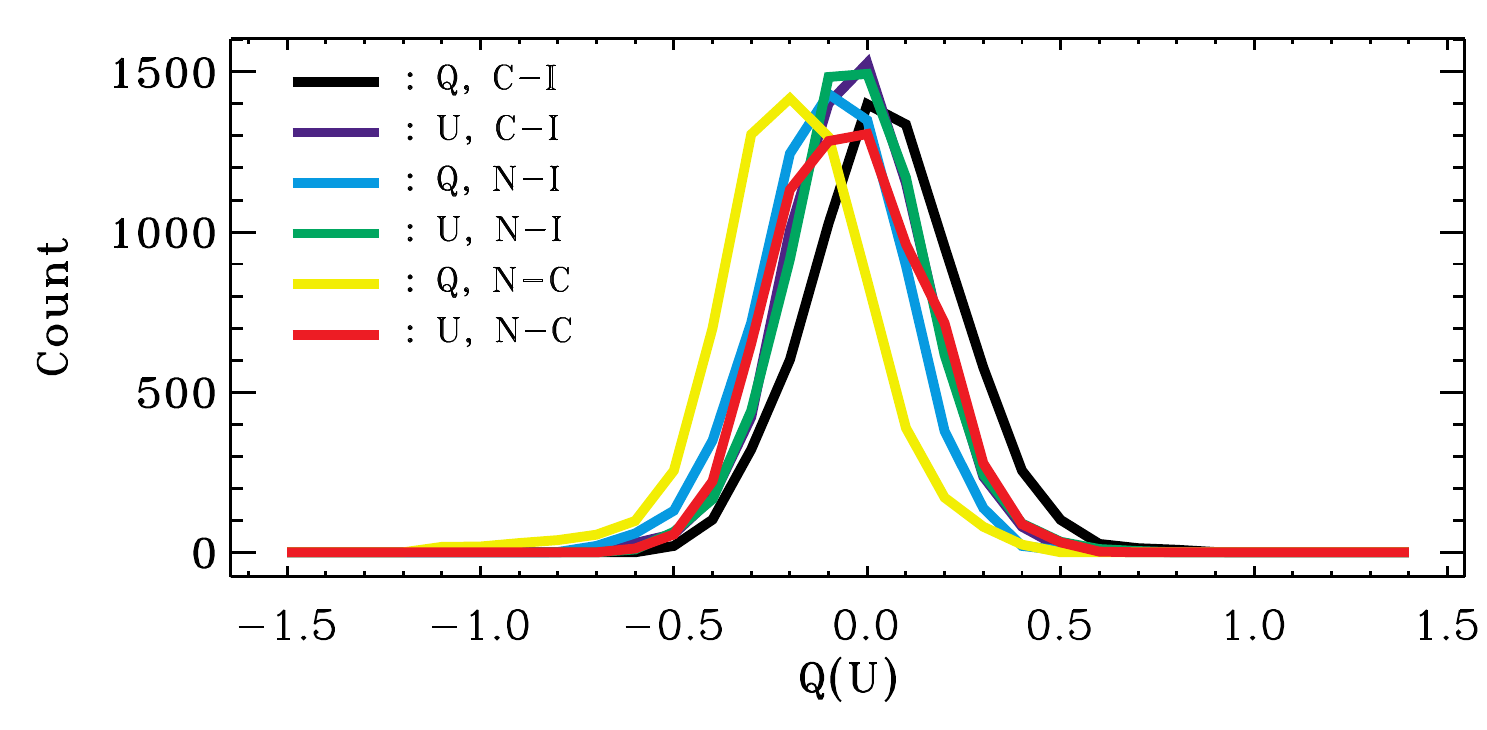}
  }
  \caption{ The differences of the Commander--ILC (left), NILC--ILC (middle) and
  Commander--NILC (right) maps in the BICEP2 zone for the Q and U Stokes
  parameters (upper-lower), and the histograms of them (bottom panel). The
  color scale corresponds to $-0.5$ to $0.5$ mK. From the top to the bottom:
  Commander--ILC, NILC--ILC and NILC--Commander. }
  \label{fig:diff N C QU}
\end{figure*}

As it is seen from table~\ref{tablea}, the deviation from
Gaussianity does not exceed $0.7\sigma$ for skewness and $1.4\sigma$ for kurtosis
for Q, and $1.8\sigma$ and $1.2\sigma$ for U for all the CMB products in the BICEP2 zone.

At first glance, based on these estimators, all the Planck 2018 CMB products are in agreement with
theoretical predictions and all the maps evenly well represent the
properties of the primordial CMB signal in the BICEP2 domain. However, we
should not forget that transition from Q and U Stokes parameters to E and B
components is characterised by very strong differences in the power spectrum
of these last components $C_E(l)$ and $C_B(l)$. As a standard prediction for
the primordial CMB we can safely adopt $C_B(l)\ll C_E(l)$ for all tensor-to-scalar ratios $r\ll 1$. That means that skewness and kurtosis test is simply
not sensitive enough for any conclusions about Gaussianity of the derived E and B-
components. In addition, we should take under consideration potential Gaussianity 
of the foregrounds \cite{Liu:2017ycx,Sebastian} in the BICEP2 domain, located far away from the Galactic plane.
Let's clarify that issue  with the following toy model.

Suppose that all the CMB products can be written as a linear combination of
the primordial signal $S_{cmb}$, the instrumental noise $n_i$, and the
residuals of systematic and foregrounds removal $S^{(i)}_r$:
\begin{eqnarray}
S_i=S_{cmb}+S^{(i)}_r+n_i
\label{eqres}
\end{eqnarray}
where index $i$ marks the SMICA, NILC, Commander, SEVEM, and ILC maps
correspondingly. 
The primordial CMB component $S_{cmb}$ obeys Gaussian statistics for all maps.
The same statistical properties can be assumed for the
instrumental noise, while for foreground residuals (see, however \cite{Liu:2017ycx,Sebastian})  and systematics we may
expect  departure from Gaussianity. From eq.~(\ref{eqres}) one can see
two potential sources of Gaussianity, $S_{cmb}$ and $n_i$. Thus, if we will
focus on the differences $M_{i,j}=S_i-S_j$, then the CMB component will not
affect the statistics of these maps. Obviously, the difference of the noises
$n_i-n_j$ will still affect the non-Gaussianity of the foregrounds and
systematics, but the CMB contribution can be greatly reduced. In the next section we will
focus on investigation of the statistical properties of the maps of
differences.

\subsection{Cross-correlations with residuals}

For estimation of possible contamination of the derived CMB products, we use
the Pearson's coefficient of cross-correlation between any two signals, $A$ and
$B$, defined as follows:
\begin{eqnarray}
&&C_{A,B}=\frac{\sum_i a_ib_i}{\sigma_A\sigma_B}\hspace{0.5cm}a_i=A_i-\langle A_i \rangle, 
b_i=B_i-\langle B_i \rangle\nonumber\\
&&\sigma_G=\left(\sum_i(G_i-\langle G_i \rangle)^2\right)^{\frac{1}{2}},\hspace{0.2cm}G=A,B
\label{pearson}
\end{eqnarray}
where $i$ indicates the pixels in the map under consideration, and
\begin{eqnarray}
\langle G_i \rangle=\sum_iG_i.
\end{eqnarray}

Suppose that signal $b$ is a linear combination of signal $a$ and some
residuals $f$, so that $b=a+f$. We are interested in the cross-correlation $C_{a,f}$
between signal $a$ and $f$, from eq.~(\ref{pearson}). Simple algebra gives us the
following expression:
\begin{eqnarray}
C_{a,f}=\left(C_{A,B}-\frac{\sigma_A}{\sigma_B}\right)\frac{\sigma_B}{\sigma_{B-A}}
\label{pearson1}
\end{eqnarray}
Here $B-A$ denotes the map of difference between the maps $B$ and $A$. Below we
will be interested in cross-correlations between the derived CMB products,
SMICA, NILC, Commander and SEVEM for Q and U components of polarisation and
the same components of the total signal at 100, 143 and 217 GHz. Thus, in
eq.~(\ref{pearson1}), the signal $a$ corresponds to CMB products, and the signal $b$
denotes the total signal for the given frequency domain. We will correlate the
corresponding Q and U components of these CMB and total signal maps to
understand how much the CMB products can be contaminated by the residuals
$f$. We will exploit the fact that if the total signal is mainly represented
by the CMB, then $C_{a,f}$ should be small.
However, if $C_{a,f}>0.2-0.3$, then the derived products are highly
contaminated by the residuals and any cosmological consequences have to be
taken with care. For practical implementation of this approach we will
consider the following model of the residuals, which can be represented as a
linear combination of the foregrounds (F), systematic effects (S) and
instrumental noise (n). Our null hypotheses  is that all CMB products (SMICA,
NILC, Commander, and SEVEM) should be characterised by almost negligible
cross-correlation coefficients in the BICEP2 zone. Below we going to present
verification of this assumption.

In the BICEP2 zone, the Pearson cross correlation coefficients between the
CMB maps and the 143 GHz map for the Q and U Stokes parameters (U in brackets)
are 0.49 (0.69), 0.54 (0.75), 0.60 (0.75), 0.52 (0.68) for the SMICA, NILC,
Commander and SEVEM maps. This level of correlations indicates that the
derived CMB products are close to the total signal at 143 GHz, but the
corresponding residuals are quite significant. In table~\ref{table1} we
show the coefficients of cross-correlations between SMICA, NILC, Commander, and
SEVEM maps with the residuals, $B-A$, for 100, 143 and 217 GHz maps in the BICEP2
zone. For the comparison, we add the corresponding coefficients for the ILC map as well.

\begin{table}[!htb]
    \centering
    \begin{tabular}{|c|c|c|c|c|c|}
        \hline
         & SMICA & NILC & Commander & SEVEM & ILC \\
        \hline
        100 GHz & $-0.08$ $(1.2)$ & $-0.34$ $(5.2)$ & $-0.21$ $(3.5)$ & $0.01$ $(0.1)$ & $-0.02$ $(0.3)$  \\
         &  $-0.10$ $(1.6)$ & $-0.34$ $(5.2)$ & $-0.22$ $(3.7)$ & $-0.09$ $(1.5)$ & $-0.08$ $(1.4)$ \\
        \hline
        143 GHz & $-0.17$ $(1.8)$ & $-0.19$ $(1.9)$ & $-0.08$ $(0.8)$ & $-0.13$ $(1.5)$ & $-0.04$ $(0.4)$  \\
        & $-0.13$ $(1.5)$ & $-0.14$ $(1.5)$ & $-0.07$ $(0.8)$ & $-0.15$ $(1.7)$ & $-0.07$ $(0.8)$ \\
        \hline
        217 GHz & $-0.23$ $(2.5)$ & $-0.23$ $(2.5)$ & $-0.16$ $(1.7)$ &  $-0.12$ $(1.3)$ & $-0.08$ $(0.8)$  \\
        & $-0.17$ $(1.8)$ & $-0.11$ $(1.2)$ & $-0.11$ $(1.2)$ & $-0.09$ $(0.9)$ & $-0.06$ $(0.7)$ \\
        \hline
    \end{tabular}
    \caption{The cross-correlation coefficients and corresponding significances (in brackets) between derived CMB products and the residuals. The upper row of each cell is for Q and the lower row for U.}
    \label{table1}
\end{table}

From table~\ref{table1} it is  seen that the ILC map is characterised by a very
low level of correlations with the residuals at all three frequency domains.
The NILC map has a relatively high level of correlations with the 100 GHz residuals,
both for Q and U. The same tendency holds for the Commander
residuals.

In order to estimate the significance of the correlations, we made the
following test. Firstly, we keep all the maps of residuals for 100--217 GHz
domain, given by the corresponding Planck 2018 CMB maps, and generate $10^3$
realisations of random Gaussian CMB, which could have  only chance
correlations with these maps. We have derived the corresponding probability
density function for those correlations, which has Gaussian character with a
variance $\sigma$. Then we divided an actual value of the correlations to the
variance. We added in table~\ref{table1} the corresponding significance of the
correlations in terms of the Gaussian standard deviations $\sigma$.

An important conclusions we can make looking at table~\ref{table1} is that, for
the 100 GHz domain, NILC and Commander maps have very significant correlations
with the residuals, while for SMICA, SEVEM and ILC they are significantly smaller.
However, moving to 217 GHz frequency domain we can see that for SMICA the
corresponding $\sigma\simeq 1.8,2.5$ and for SEVEM it is about $1.3$.
Surprisingly, our ILC map is characterised by minimal cross-correlations with
residuals for all 100--217 GHz domain and it is much more closer to
non-correlated (with residuals) Gaussian signal.

In order to understand the morphology of the contaminant of the Commander
maps, causing very significant cross-correlations with the corresponding
residuals, in figure~\ref{fig:diff N C QU} we show the Q and U maps for
differences between NILC, Commander and ILC maps in the BICEP2 domain.
\begin{table}[!htb]
 \centering
 \begin{tabular}{|c|c|c|c|c|} \hline
    Input & Q skewness & Q kurtosis & U skewness & U kurtosis \\ \hline
    Commander -- ILC  & 1.4 &  1.3 & 1.2 & 2.2 \\ \hline 
    NILC -- ILC       & 1.7 &  1.0 & 0.9 & 1.9 \\ \hline 
    NILC -- Commander & 4.3 & 10.1 & 1.1 & 1.1 \\ \hline 
\end{tabular}
 \caption{ The significances (n$\sigma$) of the skewness and kurtosis of the Q
 and U stokes parameters for NILC-ILC and Commander-ILC maps. The numbers in
 the table correspond to the parameter $n$.}
 \label{tableb}
\end{table}
For Q and U parameters, we estimated the skewness and kurtosis statistics and
found very significant departures from Gaussian expectations. We summarised
these results in table~\ref{tableb}. From figure~\ref{fig:diff N C QU} we see
that the major source of non-Gaussianity is associated with the signal
found in the right hand side bottom corner of the maps. It is not clear
whether or not it is associated with the cluster of the point-like sources with
relatively low amplitudes, not strong enough for detection by standard
methods.

\section{From Q/U to E/B maps}

In this section we would like to trace the propagation of non-Gaussian
features detected in the Q and U domain after transition to the E and B components.
The non-triviality of that propagation is associated with non-locality of
convolution and the final result potentially can be characterised by different
morphology with respect to Q/U analysis from the previous section. We will restrict
our analysis to three CMB products: NILC, Commander and ILC. Remember that in the
Q/U domain, these first two maps reveal the strongest contamination.

\begin{figure*}[!htb]
  \centering
  \includegraphics[width=0.3\textwidth]{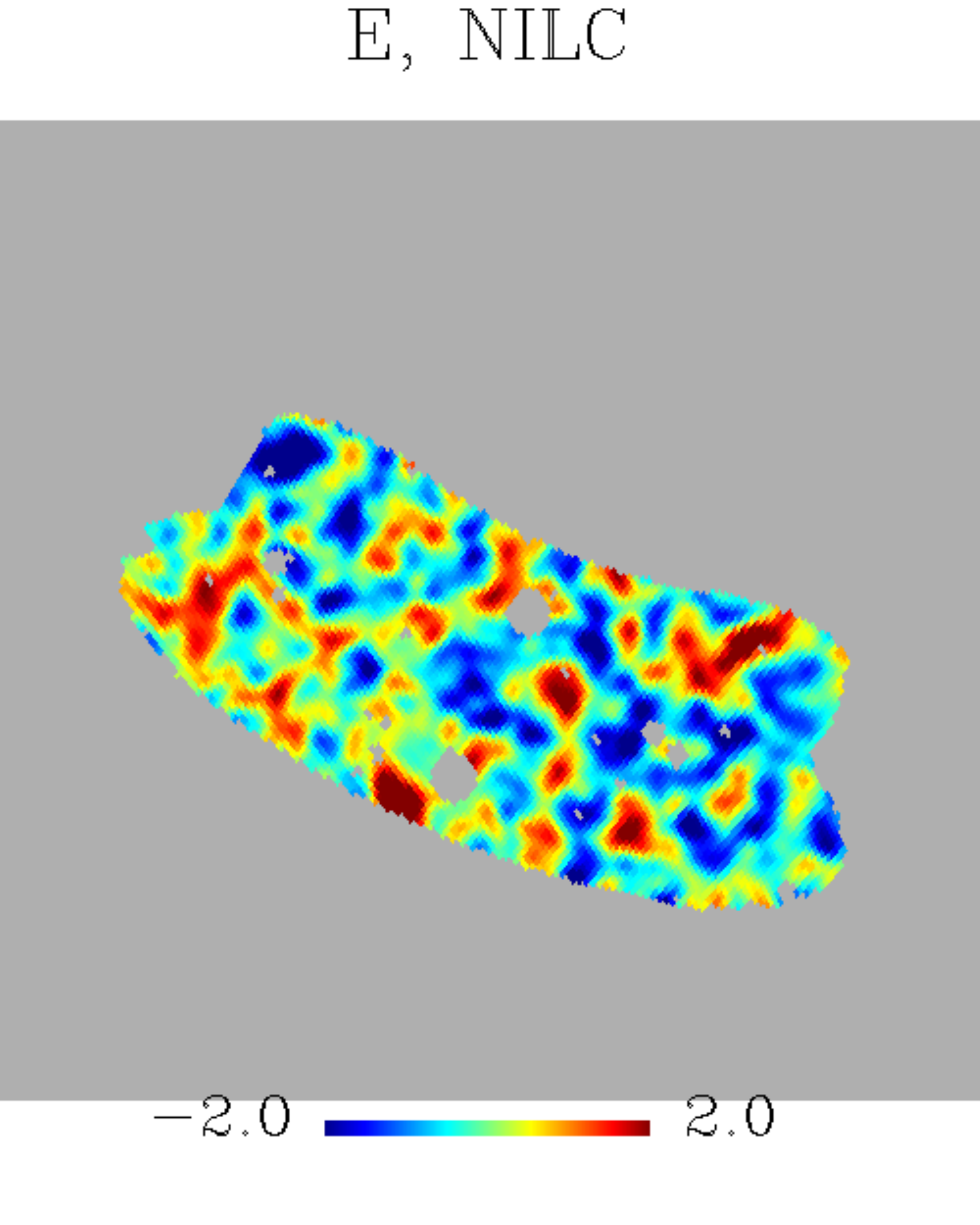}
  \includegraphics[width=0.3\textwidth]{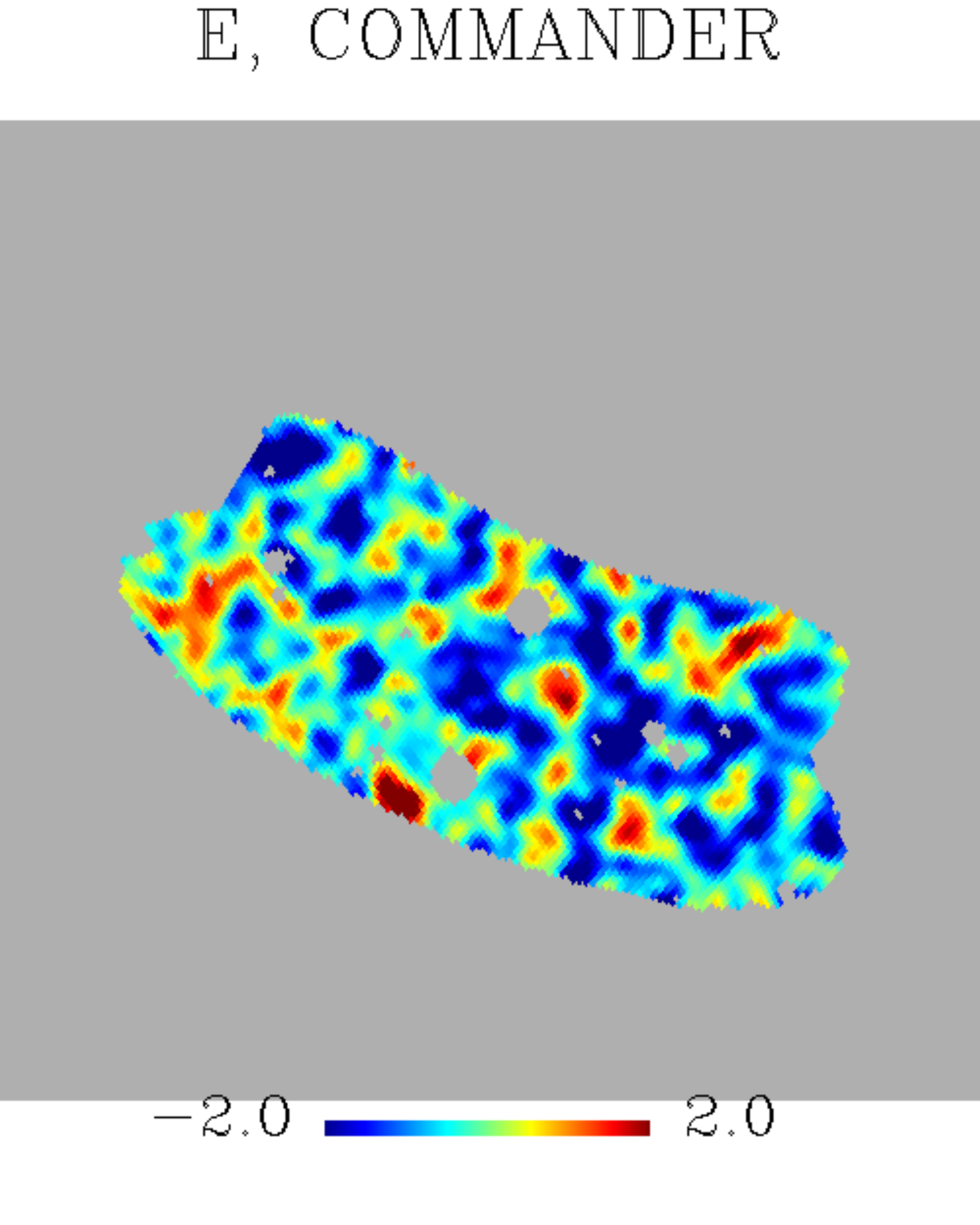}
  \includegraphics[width=0.3\textwidth]{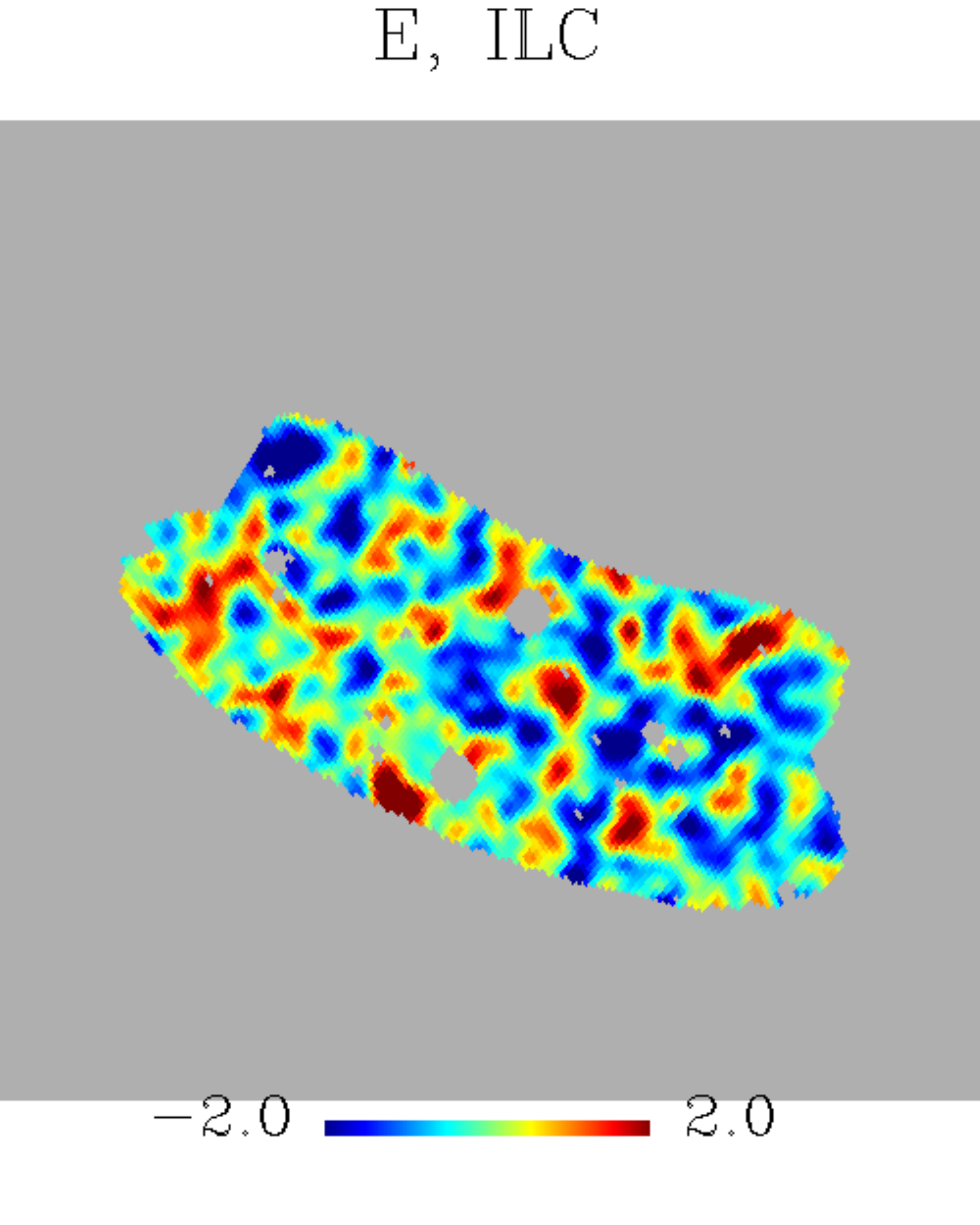}
  
   \includegraphics[width=0.3\textwidth]{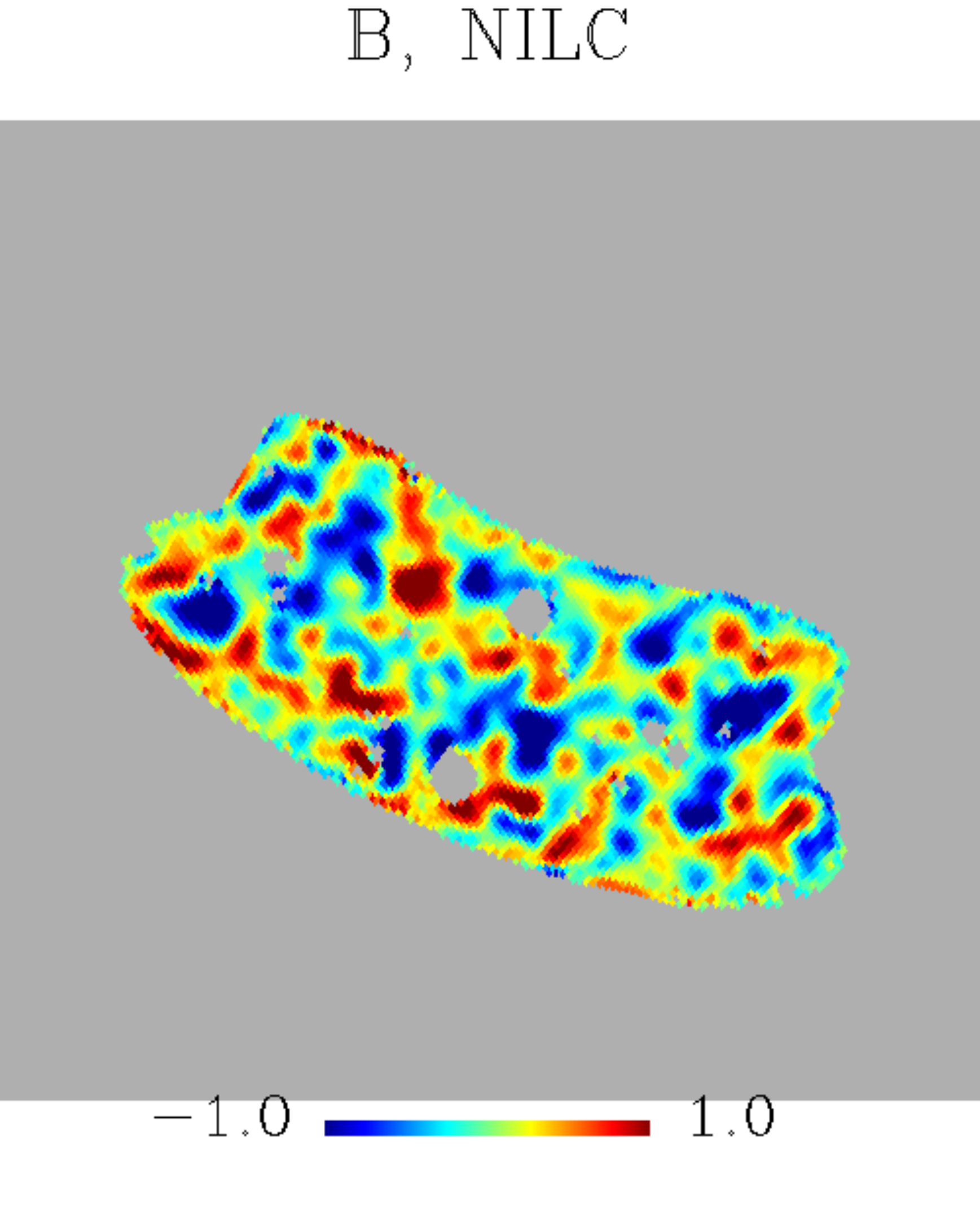}
   \includegraphics[width=0.3\textwidth]{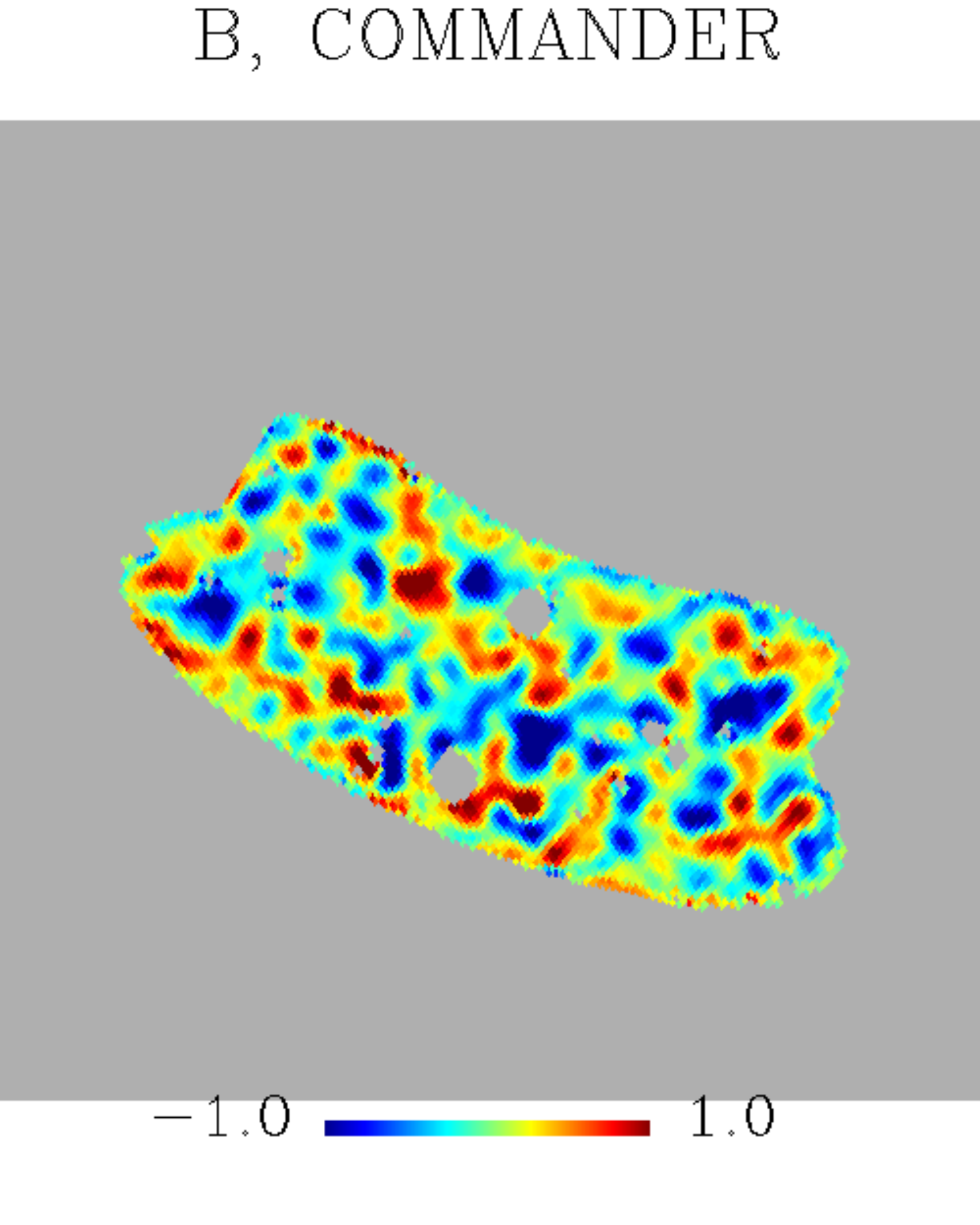}
  \includegraphics[width=0.3\textwidth]{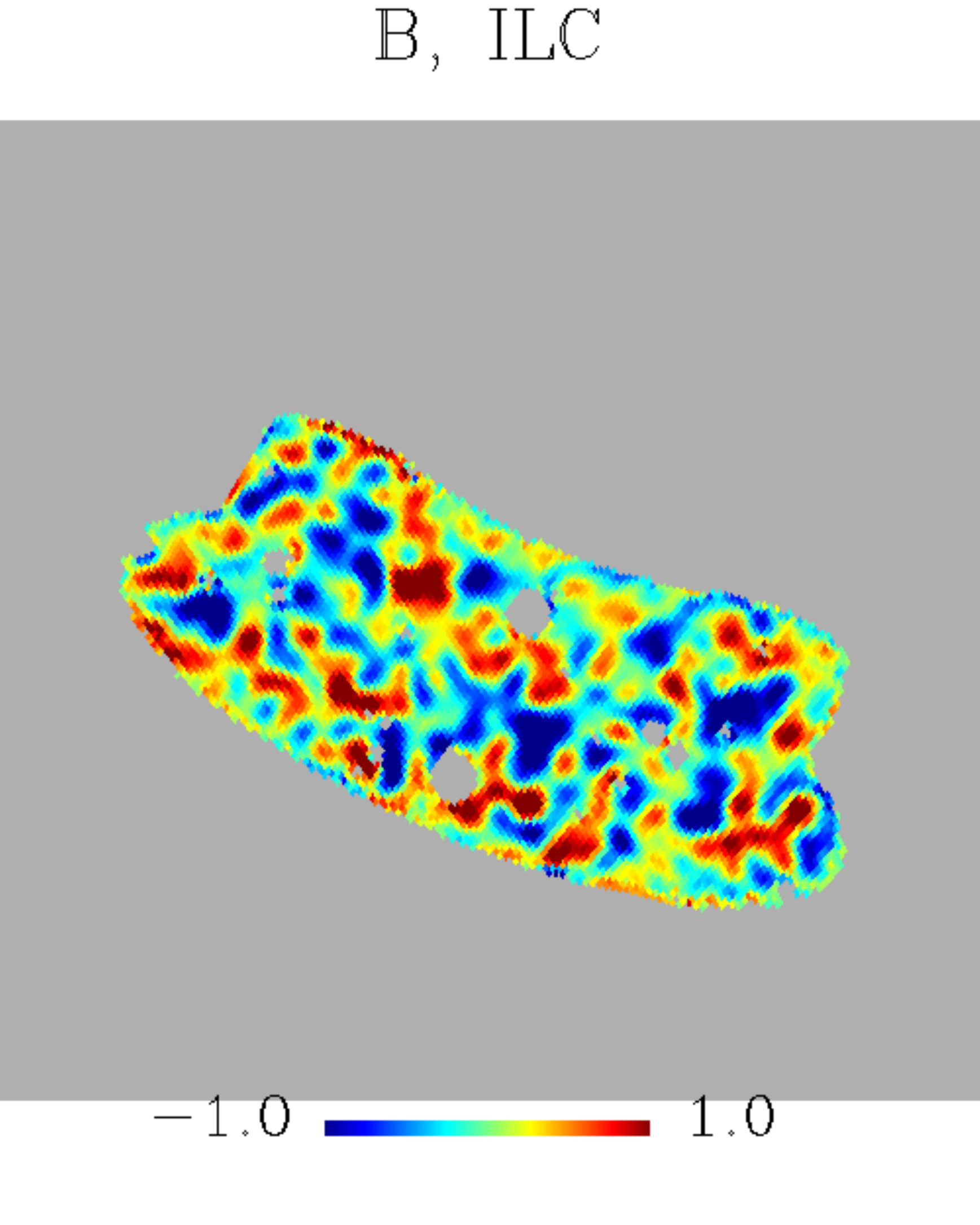} 
  
  \includegraphics[width=0.75\textwidth]{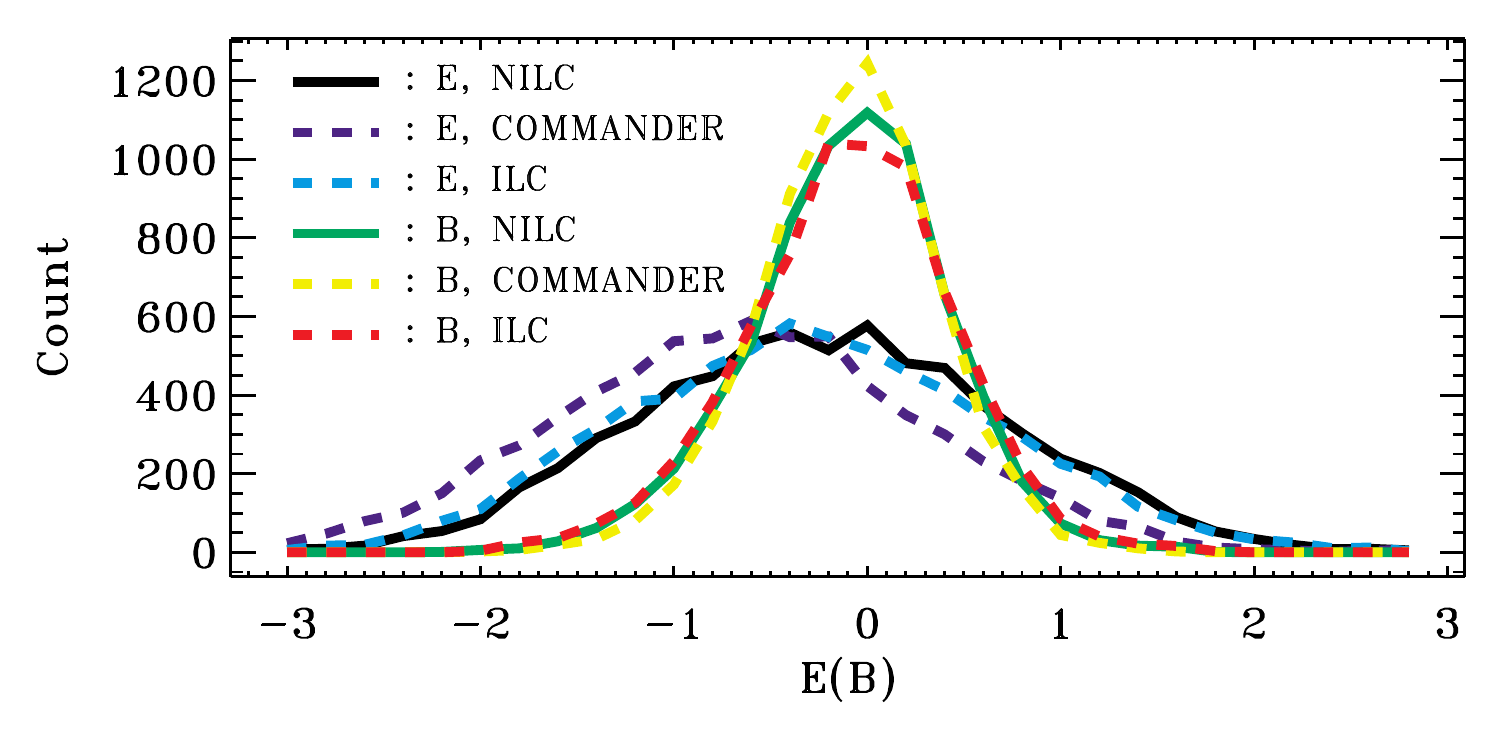}
  \caption{The E and B-components (top/middle) for the NILC, Commander and ILC
  maps. The bottom panel corresponds to the distribution function for the E and B
  components versus amplitude (mK).}
  \label{fig:cmp B}
\end{figure*}

We use the standard transformation from Q and U Stokes parameters to E and B components
with a pixel-domain EB-leakage correction.  
Note that because the input Planck maps are noisy, and the BICEP2 region is
quite small (it covers about $1\%$ of the sky), the EB-leakage is  quite
strong here and needs to removed. 
That has been done by implementation of the 
recycling methods from~\cite{2018arXiv181104691L} in the pixel domain. We show the
results  in figure~\ref{fig:cmp B}. 
The upper row of that figure represents the E-mode of polarisation, while the
bottom row corresponds to the B-mode. Very preliminary visual inspection
reveals obvious morphological similarity between NILC and ILC E-modes, and they
strongly depart from the Commander map.

From figure~\ref{fig:cmp B} one can see that, compared to the ILC and NILC maps,
the Commander E-map is strongly biased towards negative amplitudes.
For the B-mode of all the maps the distribution functions are almost identical
and Gaussian, with small fluctuations around $B=0 \ \mu$K.

In order to understand the properties of the E/B-modes, we will use  the same
model as for the Q and U Stokes parameters for each Planck 2018 product:
\begin{eqnarray}
&E_i=E_{cmb}+E^{i}_{res}+E^{i}_{noise},\nonumber\\
&B_i=B_{cmb}+B^{i}_{res}+B^{i}_{noise}
\label{eqeb}
\end{eqnarray}
where as previously, index $i$ labels the NILC, Commander and ILC maps, index
$cmb$ stands for the frequency independent CMB component, index $res$
corresponds to the residuals of the foregrounds and systematics, and $noise$
stands for instrumental noise for each Planck 2018 map, including our ILC map.
As we discussed in the previous section, the statistical properties of the
E-mode are determined by the combination of the Gaussian components ($E_{cmb},
E^i_{noise}$) and potentially the non-Gaussian tile $E_{res}$. For the B-mode,
based on \cite{2015PhRvL.114j1301B, 2015ApJ...811..126B, 2018arXiv180706211P},
we can safely neglect the CMB component, and focus on non-Gaussian residuals
and Gaussian noise.

\subsection{Analysis of the B-mode}

If the foreground residuals plus systematics do not correlate with the
instrumental noise\footnote{Note that this is a linear combination of the
noises from each frequency band.} it would be natural to assume that the shape
of the corresponding distribution function of the B-mode  will be represented
by a Gaussian distribution with the standard deviation $\sigma_B$:
\begin{eqnarray}
\sigma^2_B=\sigma^2_{noise}+\sigma^2_{res}
\label{eqeb1}
\end{eqnarray}
for all the B-mode products. Here $\sigma_{res}$ and $\sigma_{noise}$ denote the
corresponding standard deviations for the residuals and instrumental noise. If
$\sigma_{res}\ll \sigma_{noise}$, the distribution function in figure~\ref{fig:cmp B} should be similar to the PDF of the combined noise.
\begin{table}[!htb]
 \centering
 \begin{tabular}{|c|c|c|} \hline
    Input  & E & B \\ \hline
    NILC              & 0.95 & 0.46 \\ \hline
    Commander         & 0.97 & 0.51 \\ \hline
    ILC               & 0.99 & 0.53 \\ \hline
\end{tabular}
 \caption{ The standard deviation of the E- and B-modes in the BICEP2 region in $\mu$K.}
 \label{tablesd}
\end{table}

\begin{table}[!htb]
 \centering
 \begin{tabular}{|c|c|c|c|} \hline
    Maps  & Commander & 100 GHz & 143 GHz \\ \hline
    NILC  & 0.95 & 0.16 & 0.63 \\ \hline
    Commander & ---  & 0.30 & 0.69 \\ \hline
    100 GHz & ---  & ---  & 0.08 \\ \hline
\end{tabular}
 \caption{ The cross correlation between the B-mode maps of different inputs
 in the BICEP2 region. Note that the standard deviation of the 100/143 GHz
 B-maps are 0.66/0.64 $\mu K^2$, respectively.}
 \label{tablecorr}
\end{table}

 As it is seen from  figure~\ref{fig:cmp B}, for the B-mode the shape of
distribution is very close to Gaussian distribution, with the standard
deviations presented in table~\ref{tablesd}. These standard deviations
differ between maps by about $10-20\%$, and the major sources of that
difference are found in the signals with amplitudes $B < \sigma_B$, while for
$B>\sigma_B$ the corresponding functions are remarkably similar. One can trace
this similarity from figure~\ref{fig:cmp B} (middle panel), looking at the
brightest positive and negative peaks. In order to understand the origin of
that structure, one should look at the cross-correlation coefficients between
some of the CMB products and the total B-mode signal at 100 and 143 GHz,
presented in table~\ref{tablecorr}. Firstly, NILC and Commander B-mode signals
are correlated at 0.95 level. Secondly, the NILC B-mode consist at the level
of 63\% with 143 GHz signal and 16\% with 100 GHz map. For the Commander map
these coefficients are 69\% and 30\% correspondingly.

Strong correlations of the CMB products with the total frequency maps can be
understood in terms of projection of the CMB products to the 100 and 143 GHz maps.
For Commander, the amplitude of the B-mode can be presented as:
\begin{eqnarray}
B_c=\alpha B_{100}+\beta B_{143}
\label{b1}
\end{eqnarray}
where the coefficients $\alpha$ and $\beta$ are related to the
cross-correlations coefficients $C^c_{100},C^c_{143}$ in table~\ref{tablecorr}
as:
\begin{eqnarray}
\alpha=\frac{1}{D}\left(dC^c_{100}-bC^c_{143}\right),\hspace{0.2cm}
\beta=\frac{1}{D}\left(aC^c_{143}-cC^c_{100}\right)
\label{b2}
\end{eqnarray}
and
\begin{eqnarray}
D=ad-bc,\hspace{0.2cm}a=\frac{\sigma_{100}}{\sigma_c},\hspace{0.2cm}
b=C_{100,143}\frac{\sigma_{143}}{\sigma_{100}},\hspace{0.2cm}
c=C_{100,143}\frac{\sigma_{100}}{\sigma_{c}},\hspace{0.2cm}
d=\frac{\sigma_{143}}{\sigma_c}.
\label{b3}
\end{eqnarray}
Here $C_{100,143}$ and $C^c_{100}$,$C^c_{143}$ are the cross-correlation
coefficients between 100 and 143 GHz maps, and between Commander and 100,143
GHz maps correspondingly.

After simple algebra, from eqs.~(\ref{b2}-\ref{b3}) and table~\ref{tablecorr} one
can get: $\alpha\simeq 0.20$ and $\beta\simeq 0.53$. Thus, the Commander map
consists of $\simeq 53\%$ of the signal from 143 GHz map, and $\simeq 20\%$ of
the signal from 100 GHz map. The corresponding cross-correlations for Commander
and NILC are very high for 143 GHz. They are almost 8 times greater then between
100 and 143 GHz.

\subsection{Analysis of the E-mode}

The E-mode of polarisation has a very important difference witn respect to the
B-mode. In addition to the noise and residuals, the E-mode contains a significant primordial
CMB component. This is why the distribution function for the E-mode in
figure~\ref{fig:cmp B} has a standard deviation almost two times
greater than the B-mode (see table~\ref{tablesd}). The common feature for all
E-mode CMB products is a shift of the distribution to the negative amplitudes
(mean values) by 0.3--0.7 $\mu K$, shown in  figure~\ref{fig:cmp B}.

In section 2.2 we have investigated the cross-correlations between the Stokes
parameters Q and U for NILC and Commander and the corresponding residuals at
100 and 143 GHz. These cross-correlations are about $3.5\sigma$ to $5\sigma$ away from
non-correlated Gaussian realisations. Here, in table \ref{tableE1}, we  present
the correlations of the E-mode with the residuals, in order to see the
effect of propagation of the Q/U anomalies to the E-mode.
\begin{table}[!htb]
 \centering
 \begin{tabular}{|c|c|c|c|c|c|} \hline
    Input & C & 100 - N & 100 - C & 143 - N & 143 - C \\ \hline
    NILC (N)         &  $0.99$  & $-0.25$ & $-0.24$ $(3.8)$ & $-0.21$ & $-0.18$ $(2.8)$ \\ \hline 
    Commander (C)    &   --   & $-0.18$ $(2.8)$ & $-0.19$ & $-0.17$ $(2.7)$ & $-0.17$ \\ \hline 
    100 - N          &   --   &  --   & $0.98$  & $0.12$  & $0.01$ \\ \hline
    100 - C          &   --   &  --   &  --   & $0.07$  & $0.00$ \\ \hline 
    143 - N          &   --   &  --   &  --   &  --   & $0.97$ \\ \hline 
    143 - C          &   --   &  --   &  --   &  --   &  --  \\ \hline
\end{tabular}
 \caption{ The cross-correlation coefficients for NILC, Commander and the
 corresponding  E-mode residuals maps at 100 and 143 GHz, and the
 significances (in brackets). Note that the significances can only be computed
 when at least one side is the CMB map.}
 \label{tableE1}
\end{table}

\section{Skewness and kurtosis for E and B modes}

In the previous section we discussed the properties of the E- and B-modes in
connection with their cross-correlations with the residuals at 100 and 143 GHz.
Here we will investigate the third and forth moments of distribution functions
of the E- and B-modes.

We have shown that due to non-locality of transition from Q/U components to E/B,
the anomalies of the distribution function and skewness and kurtosis are
significantly diluted. At the same time, the E-mode for the CMB maps is
systematically shifted to the negative amplitudes, while the B-mode has
a discrepancy at very low amplitudes. For the B-mode we have detected very
significant cross-correlations between NILC and Commander maps with 143 GHz
total map. Taking into account that for the Planck data we can safely
neglect the primordial signal, these correlations can be related to absorption
of the 143 GHz noise by the B-mode.
 \begin{figure*}[!htb]
  \centering
 \includegraphics[width=0.32\textwidth]{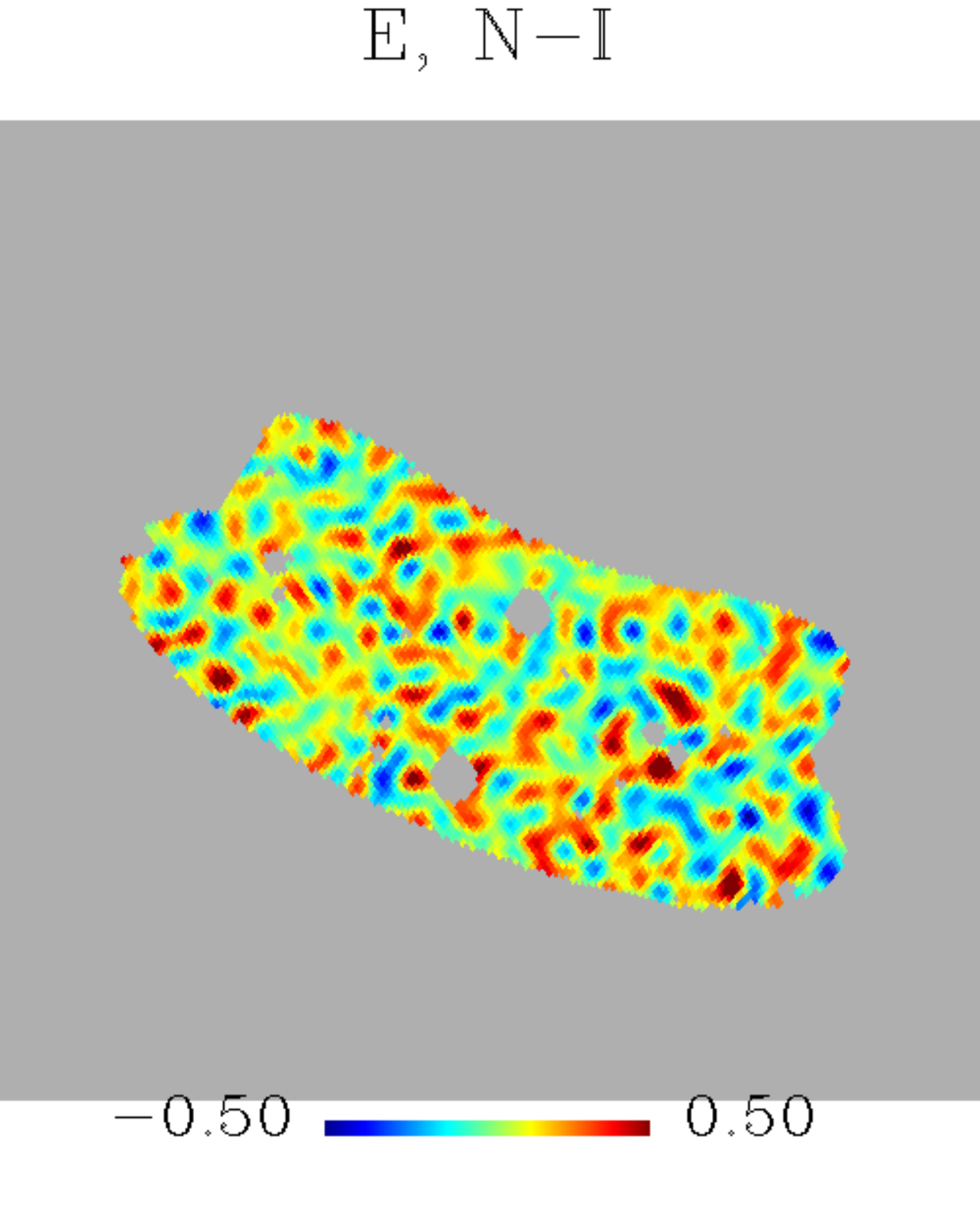}
 \includegraphics[width=0.32\textwidth]{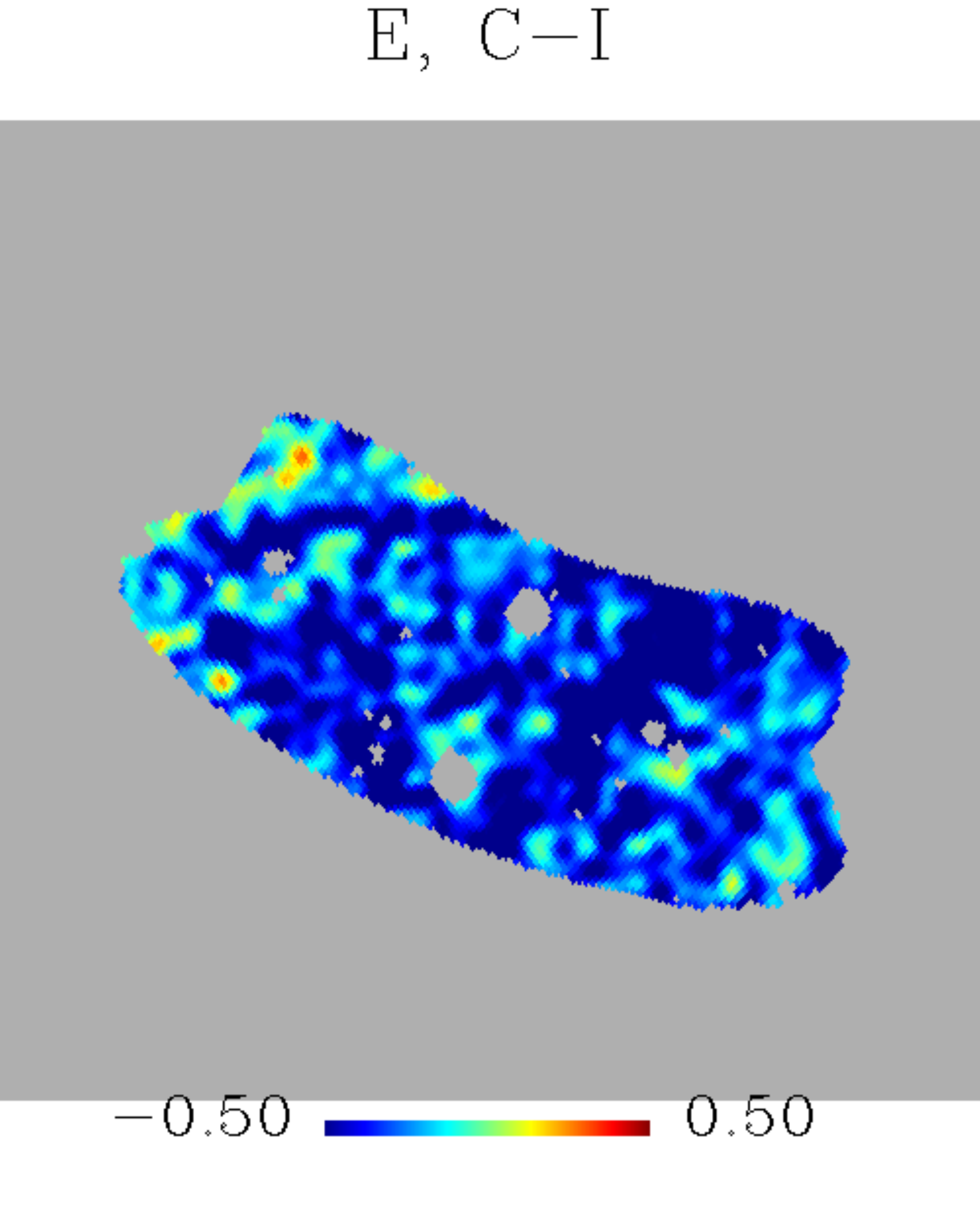}
 \includegraphics[width=0.32\textwidth]{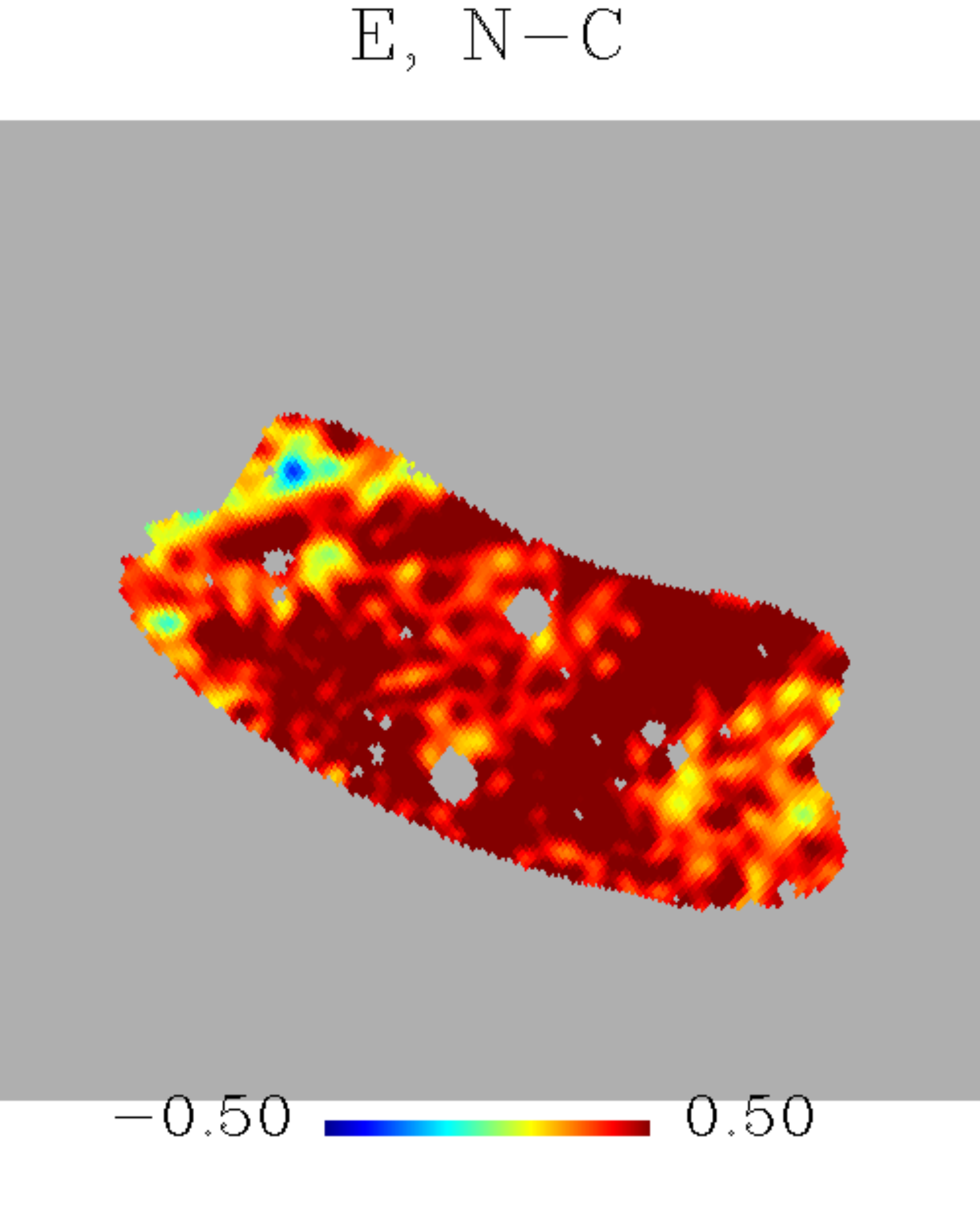}
 
  \includegraphics[width=0.32\textwidth]{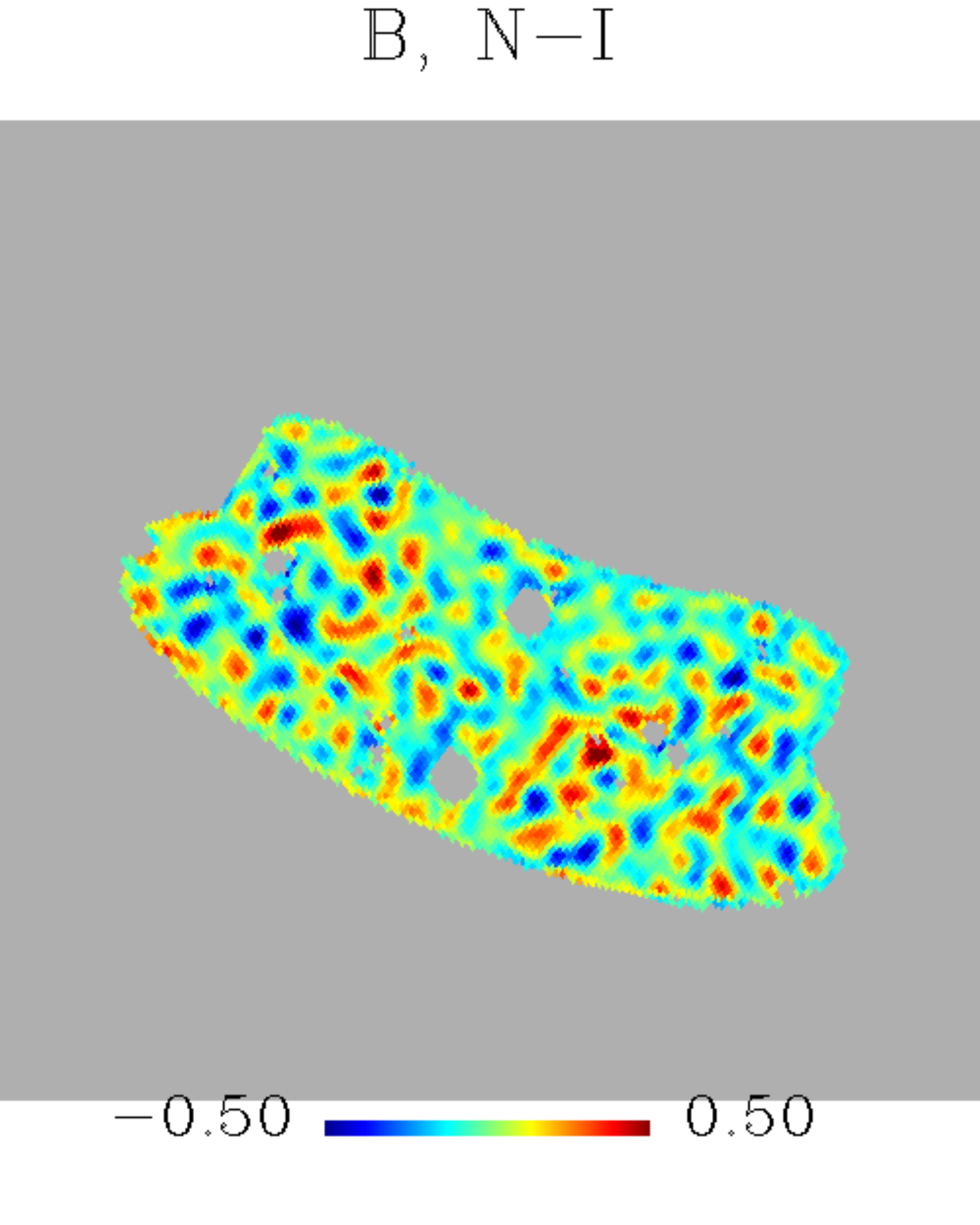}
  \includegraphics[width=0.32\textwidth]{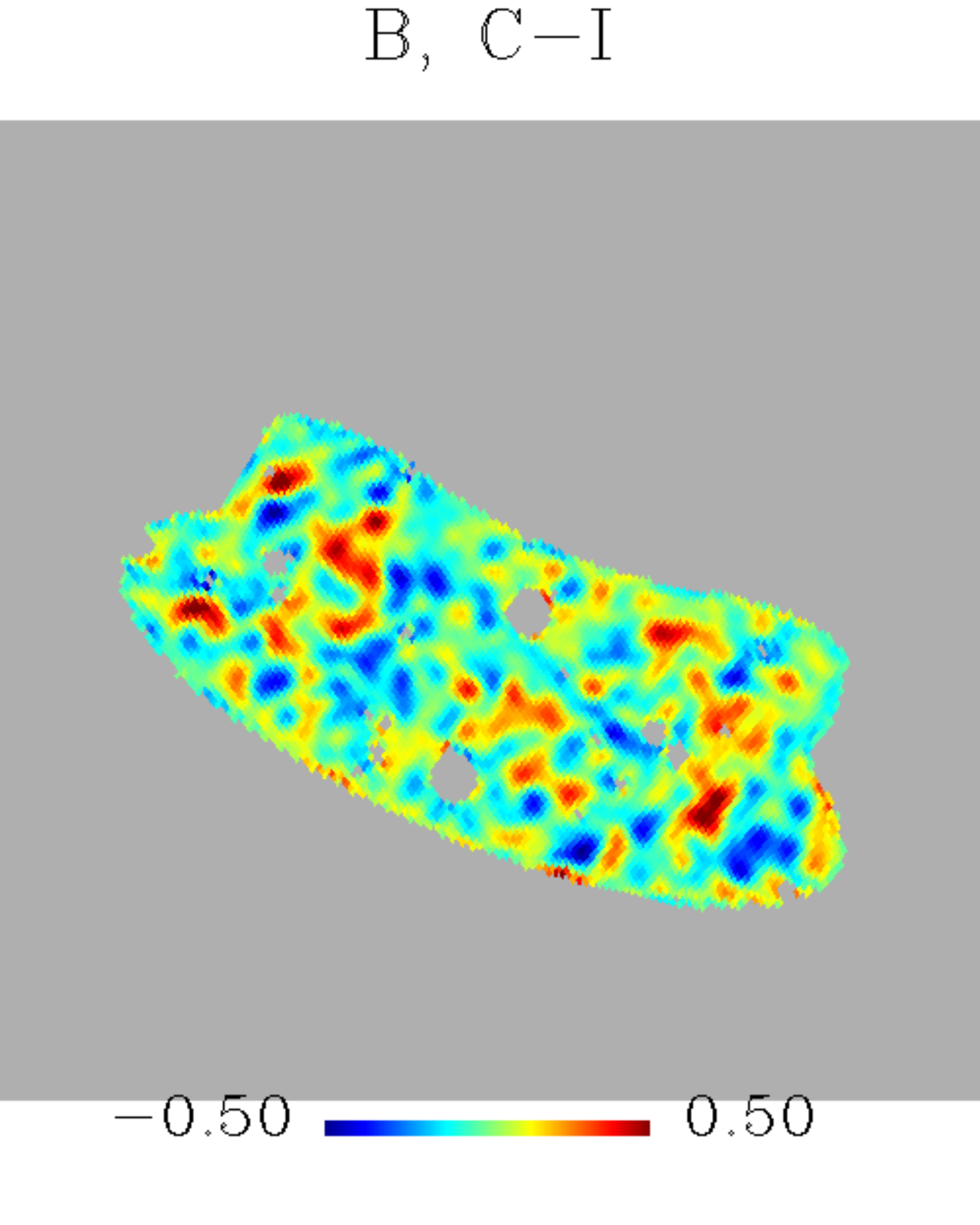}
  \includegraphics[width=0.32\textwidth]{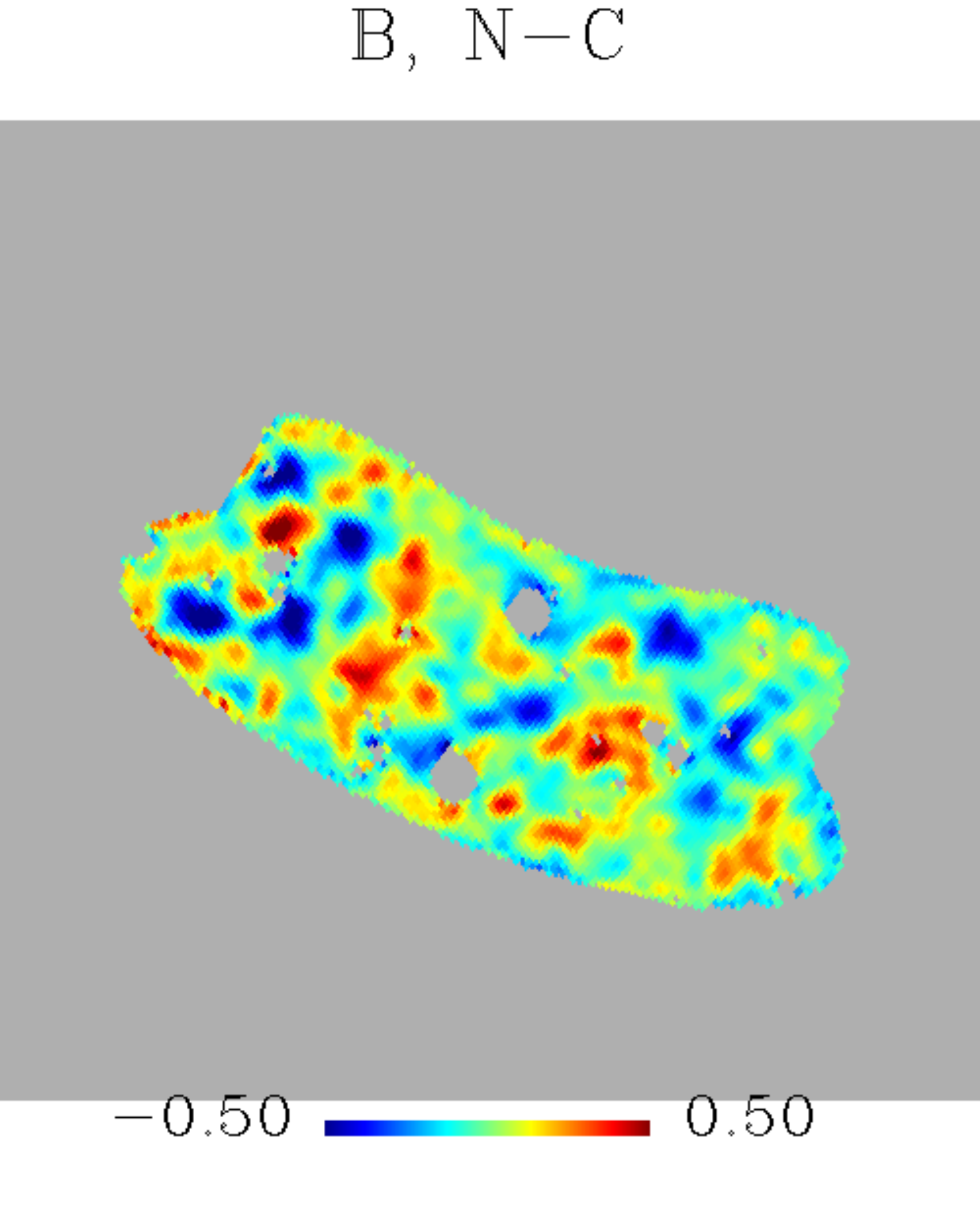}
  \caption{The maps of differences for E (upper row) and B (lower row)
  components for NILC--ILC (left), Commander--ILC (middle) and NILC--Commander
  (right panel).}
  \label{fig:cmp Bdif}
\end{figure*}

\begin{figure*}[!htb]
  \centering
 \includegraphics[width=0.8\textwidth]{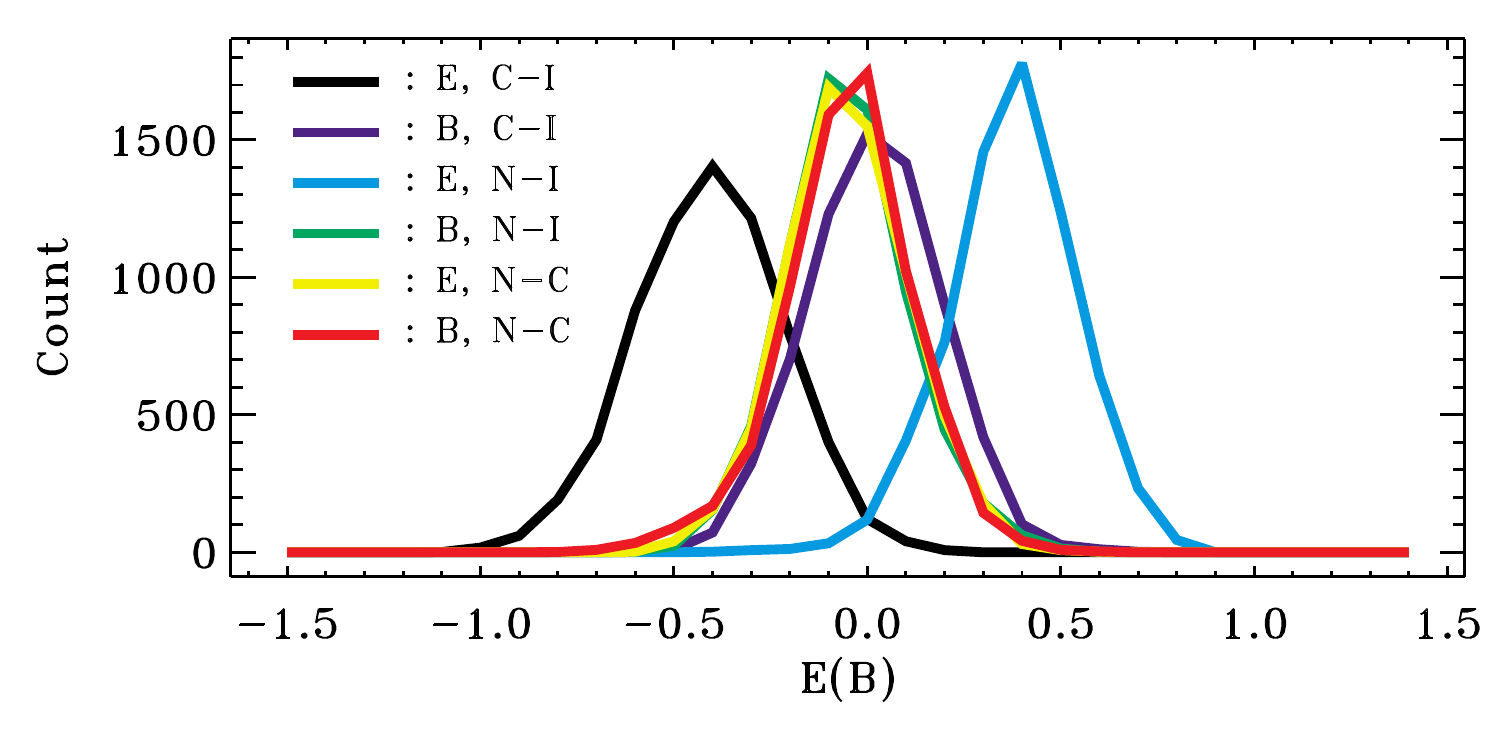}
  \caption{Histograms of the E- and B-maps of differences for Commander, NILC and ILC CMB products.}
  \label{fig:cmp Bdif1}
\end{figure*}

From  figure~\ref{fig:cmp B} we have seen that the distribution function for the
B-mode is close to Gaussian, while the E-mode is characterised by a
relatively strong shift to negative mean values. However, it is not the only
feature of the E-mode distribution. In table~\ref{tableb2} we show the skewness
and kurtosis of the E- and B-modes for the NILC, Commander and ILC polarisation maps,
and of their differences, presented in figure~\ref{fig:cmp Bdif} and
figure~\ref{fig:cmp Bdif1}.

\begin{table}[!htb]
 \centering
 \begin{tabular}{|c|c|c|c|c|} \hline
    Input & E-skewness & E-kurtosis & B-skewness & B-kurtosis \\ \hline
    NILC             & 1.3 &  2.6 & 1.7 & 1.2 \\ \hline 
    Commander        & 1.5 &  2.0 & 1.5 & 1.1 \\ \hline 
    ILC              & 1.9 &  2.9 & 1.5 & 0.8 \\ \hline\hline                                         
    Commander--ILC  & 0.5 &  0.2 & 1.7 & 0.4 \\ \hline 
    NILC--ILC       & 0.3 &  0.1 & 0.6 & 0.3 \\ \hline 
    NILC--Commander & 2.5 &  2.7 & 2.3 & 2.6 \\ \hline
\end{tabular}
 \caption{ The significance ($n\sigma$) of the skewness and kurtosis of E and
 B maps NILC, Commander, ILC maps and their differences. The numbers in the
 table correspond to the parameter $n$.}
 \label{tableb2}
\end{table}
The E-modes for NILC, Commander and ILC are characterised by $2.0\sigma$ to $2.9\sigma$
departures from Gaussian realisations for the kurtosis. Note that the E- and B-mode
for the difference NILC--Commander are peculiar at the level of
$2.3\sigma$ to $2.7\sigma$ for the skewness and kurtosis. The comparison of the
results, presented in table~\ref{tableb2} and table~\ref{tableb} shows that
the significance of the skewness and kurtosis under transition from Q/U to E/B is
significantly reduced. This effect is quite understandable. This transition
has non-local character as in multipole, as in pixel domain. Due to
non-locality, the non-Gaussian features are redistributed among different
pixels, reducing peculiarity of the E/B maps. Thus, one of the major
conclusions following from our analysis is that tests on non-Gaussianity of E/B-
map without Q/U components can easily produce misleading effects.

\section{Conclusion}

In our paper we have investigated the statistical properties of polarisation
in the BICEP2 zone, basing our analysis on  the Planck 2018 CMB products. One
of the most interesting conclusions from our analysis is that Q and U Stokes
parameters are more sensitive to the foreground residuals and possible
systematics than the derived E/B-modes. We believe this effect is mainly due to
non-locality of the $Q/U\rightarrow E/B$ transition, and removal of the E/B
leakage. These tests have not been used previously in the analysis of
statistical isotropy and non-Gaussianity of the Planck 2018 CMB
products~\cite{2016A&A...594A..16P, 2019arXiv190602552P} and allow us to evaluate
the effectiveness of various methods for foreground component separation and
validation of the polarization CMB signals.

We have shown that the NILC and Commander maps have very strong correlations with the residuals at 100 GHz with significance $3.5-5.2\sigma$, both for Q and U Stokes
parameters.

After transition to the E/B modes, we have found a $2.6\sigma$ departure of the
E-mode kurtosis statistic from Gaussianity for the NILC map, while the Commander map
is characterised by a departure at the $2\sigma$ level. For these maps the B-mode skewness and
kurtosis statistics lies within $1.5-1.7\sigma$. We have extended our analysis
for the map of difference between NILC and Commander E/B-maps, in order to
test the hypothesis that both these maps contain sub-dominant non-Gaussian
residuals. We have found that E/B skewness and kurtosis for the difference map are peculiar at the level of $2.3-2.7\sigma$.

In conclusion, we would like to emphasize that the results obtained from the estimators adopted in this paper demonstrate that simple testing of statistical isotropy and non-Gaussian polarization of E and B modes
is insufficient to confirm their cosmological nature
for the new generation of CMB experiments. It is therefore important to complement
these tests by analyzing the Q and U components, and especially their cross-correlations with the residual signals for a wide range of frequencies. When considering the potential
Gaussian foregrounds far from the galactic plane, the analysis of these cross-correlations may be the most informative test.


\Ack{

This research has made use of the
\textsc{HEALPix}~\citep{2005ApJ...622..759G} package
and was partially funded by the 
Villum Fonden through
the Deep Space project. Hao Liu is also supported by the National Natural
Science Foundation of China (Grants No. 11653002, 11653003), the Strategic
Priority Research Program of the CAS (Grant No. XDB23020000) and the Youth
Innovation Promotion Association, CAS. \\

}


\bibliography{newbib}

\end{document}